\documentclass[%
reprint,
superscriptaddress,
amsmath,amssymb,
aps,
pra,
]{revtex4-2}
\usepackage{xcolor}
\usepackage{graphicx}
\usepackage{dcolumn}
\usepackage{bm}
\usepackage[hidelinks]{hyperref} 

\begin{document}
	
	\title{Experimental Demonstration of an On-Axis Laser Ranging Interferometer for Future Gravity Missions}
	
	\author{Daikang Wei}
	\email{daikang.wei@aei.mpg.de}
	\author{Christoph Bode}%
	\affiliation{%
		Max-Planck-Institut f\"ur Gravitationsphysik (Albert-Einstein-Institut), Callinstraße 38, 30167 Hannover, Germany
	}%
	\affiliation{%
		Leibniz Universit\"at Hannover, Institut für Gravitationsphysik, Callinstraße 38, Hannover 30167, Germany
	}%
	\author{Kohei Yamamoto}%
	\affiliation{%
		Center for Space Sciences and Technology, University of Maryland, Baltimore County, 1000 Hilltop Circle, Baltimore, MD 21250, USA
	}%
	\affiliation{%
		Gravitational Astrophysics Lab, NASA/GSFC, 8800 Greenbelt Road, Greenbelt, MD 20771, USA
	}%
	\affiliation{%
		Center for Research and Exploration in Space Science and Technology, NASA/GSFC, 8800 Greenbelt Road, Greenbelt, MD 20771, USA
	}%
	\author{Yongho Lee}%
	\author{Germ\'an Fern\'andez Barranco}%
	\author{Vitali M\"{u}ller}%
	\affiliation{%
		Max-Planck-Institut f\"ur Gravitationsphysik (Albert-Einstein-Institut), Callinstraße 38, 30167 Hannover, Germany
	}%
	\affiliation{%
		Leibniz Universit\"at Hannover, Institut für Gravitationsphysik, Callinstraße 38, Hannover 30167, Germany
	}%
	\author{Miguel Dovale \'Alvarez}%
	\affiliation{%
		Wyant College of Optical Sciences, University of Arizona, 1630 E. University Blvd., Tucson, AZ 85721, USA
	}%
	\author{Juan Jos\'e Esteban Delgado}%
	\author{Gerhard Heinzel}%
	\affiliation{%
		Max-Planck-Institut f\"ur Gravitationsphysik (Albert-Einstein-Institut), Callinstraße 38, 30167 Hannover, Germany
	}%
	\affiliation{%
		Leibniz Universit\"at Hannover, Institut für Gravitationsphysik, Callinstraße 38, Hannover 30167, Germany
	}%
	
	\date{\today}
	
	\begin{abstract}
		We experimentally demonstrate a novel interferometric architecture for next-generation gravity missions, featuring a laser ranging interferometer (LRI) that enables monoaxial transmission and reception of laser beams between two optical benches with a heterodyne frequency of 7.3 MHz. Active beam steering loops, utilizing differential wavefront sensing (DWS) signals, ensure co-alignment between the receiving (RX) beam  and the transmitting (TX) beam. With spacecraft attitude jitter simulated by hexapod-driven rotations, the interferometric link achieves a pointing stability below 10 µrad/$\mathrm{\sqrt{Hz}}$ in the frequency range between 0.2 mHz and 0.5 Hz, and the fluctuation of the TX beam's polarization state results in a reduction of  0.14\% in the carrier-to-noise-density ratio over a 15-hour continuous measurement. Additionally, tilt-to-length (TTL) coupling is experimentally investigated using the periodic scanning of the hexapod. Experimental results show that the on-axis LRI enables the inter-spacecraft ranging measurements with nanometer accuracy, making it a potential candidate for future GRACE-like missions.
	\end{abstract}
	
	\maketitle

	\section{INTRODUCTION}
	
	The Gravity Recovery and Climate Experiment (GRACE) mission \cite{tapley2004gravity}, a collaboration between the National Aeronautics and Space Administration (NASA) and the German Aerospace Center (DLR), was operational from March 2002 to October 2017. It employed twin satellites separated by approximately 200 kilometers to precisely measure inter-spacecraft distance variations via a Microwave Instrument (MWI), enabling the retrieval of Earth’s time-variable gravity field \cite{tapley2005ggm02}. Monthly gravity field maps derived from the GRACE data were used to investigate global-scale mass transport phenomena, including ice-sheet dynamics, ocean circulation patterns, and continental hydrologic processes, which are critical for understanding the dynamics of Earth’s climate system \cite{tapley2004grace}.
	
	As the successor to the GRACE mission, the GRACE Follow-On (GRACE-FO) mission was launched in May 2018 to extend continuous monitoring of Earth’s gravity field \cite{Kornfeld2019}. In addition to the MWI serving as the primary instrument, the mission incorporated a Laser Ranging Interferometer (LRI) as a technology demonstration. To mitigate temporal gaps between the GRACE-FO mission and its successors, future GRACE-like missions are under active development. GRACE-Continuity (GRACE-C), a joint mission between the United States and Germany, is scheduled for launch in 2028 to ensure uninterrupted data collection \cite{bender2025short}. In parallel, the European Space Agency (ESA) is progressing with the Next-Generation Gravity Mission (NGGM), which will employ a Bender constellation to enhance spatial and temporal resolution \cite{haagmans2020esa}, with an anticipated launch in the early 2030s.
	
	The GRACE-FO LRI successfully tracked the inter-spacecraft distance with a sensitivity level of 1 nm/$\mathrm{\sqrt{Hz}}$ at frequencies around 100 mHz, which is two orders of magnitude better than the performance of the MWI \cite{ghobadi2020grace,abich2019orbit}. The GRACE-FO LRI utilizes optical heterodyne interferometry in a transponder configuration, representing the first demonstration in space of inter-spacecraft laser links \cite{sheard2012intersatellite}. In the reference spacecraft, a laser is frequency-stabilized to a highly stable optical cavity. A beamsplitter diverts a small fraction of the laser to serve as a local oscillator beam, while most of the laser is transmitted to the remote spacecraft. In the transponder spacecraft, the laser is frequency-offset-locked to the incoming beam and transmitted back to the reference spacecraft \cite{abich2019orbit,heinzel2006lisa}. The returning beam is then combined with the local beam to produce the heterodyne signal, enabling the measurement of relative phase shifts that encode the round-trip distance variation between the two spacecraft. The reference and transponder spacecraft are built identically, allowing them to switch operational roles. In addition, an active beam steering method was implemented in the LRI to mitigate the effect of the spacecraft attitude jitter \cite{schutze2014laser}. Owing to the success of the GRACE-FO LRI, the LRI is selected to be the primary instrument in the upcoming GRACE-C and NGGM missions \cite{landerer2024towards,nicklaus2022towards,nicklaus2020laser}.

	The GRACE-FO LRI employs an off-axis configuration using a triple mirror assembly (TMA), which introduces a lateral offset between the receiving (RX) and transmitting (TX) beams \cite{schutze_retroreflector_2014,ward2014design}. This design circumvents the constraint imposed by the MWI, which occupies the line of sight between the centers of mass of the two spacecraft. To suppress the tilt-to-length (TTL) coupling that arises from the spacecraft attitude jitter, the virtual vertex of the TMA is aligned with the center of mass (CoM) of the spacecraft, where the accelerometer is located. The LRI will be the main instrument in future missions. Without the accommodation constraints imposed by the MWI, it becomes feasible to investigate different topologies. Mandel et al. designed a mono-axis architecture using a retro-reflector and polarizing optics, which integrates multiple methodologies for heterodyne laser ranging while enabling beam-tracking capabilities \cite{mandel2020architecture}. Our group has proposed an on-axis architecture for laser ranging systems, which utilizes polarization-dependent routing to maintain the anti-parallelism between the RX beam and the TX beam. This design was thoroughly validated through numerical simulations, demonstrating its potential for high-performance laser ranging \cite{yang2022axis}. The experimental result presented here extends this framework by incorporating modifications to the proposed on-axis topology.
	
	This paper reports the experimental results of an on-axis LRI, which establishes a transponder-based laser interferometric link featuring two independent active beam steering loops. A laboratory-developed breadboard model demonstrates an on-axis interferometric link via anti-parallel RX and TX  beams between two optical benches. With the angular jitters, the beam pointing stability was characterized using calibrated differential wavefront sensing (DWS) and differential power sensing (DPS) signals. The potential fluctuation of the TX beam's polarization state was quantified, and its impact on the carrier-to-noise-density ratio ($\mathrm{C/N_0}$), defined as the ratio of carrier power to noise power spectral density, was analyzed. Additionally, the TTL coupling of the optical bench was investigated through dedicated rotational configurations based on the experimental model.

    Beyond the GRACE-like missions, the concepts and technologies demonstrated in this work could have broader application potential. In the field of gravitational wave detection, the Laser Interferometer Space Antenna (LISA) mission will establish three laser interferometer links between three spacecraft arranged in a triangular array \cite{LISA}. Related space-based initiatives include the Chinese Taiji \cite{Luo2020} and Tianqin \cite{Luo2016} projects, as well as the Japanese DECi-hertz Interferometer Gravitational wave Observatory (DECIGO) \cite{kawamura2011japanese}. Additionally, the precision beam steering and DWS technologies developed here are applicable to laser communication systems, where maintaining precise alignment is critical for minimizing link loss and maximizing data throughput. Moreover, the space interferometric technology explored here could facilitate fundamental experiments in space quantum interference \cite{vallone2016interference, wu2024single, lu2022micius}.

	The paper is structured as follows: Section \ref{concept} introduces the concept of the on-axis LRI. The experimental setup and methodology are detailed in Section \ref{methods}, which provides a comprehensive overview of the on-axis laser link, as well as the implementation of DWS and DPS technologies. Section \ref{results} presents the experimental results, focusing on angle calibration, beam pointing, polarization effects, and TTL coupling. Finally, Section \ref{summary} provides a summary and concluding remarks.

	\section{ON-AXIS LRI CONCEPT}\label{concept}
	
	\begin{figure*}
		\centering
		\includegraphics[width=0.95\textwidth]{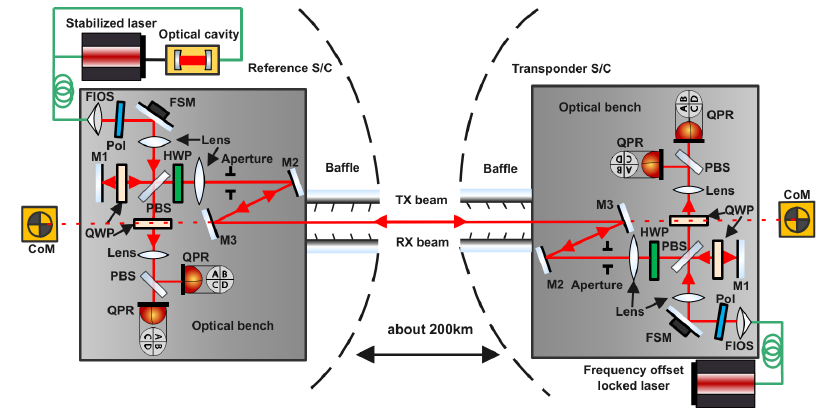}
		\caption{Concept sketch of the inter-spacecraft laser interferometry in on-axis configuration. S/C, spacecraft; FIOS, ﬁber-injector optical subassembly; Pol, polarizer; FSM, fast steering mirror; M, mirror; QWP, quarter-wave plate; HWP, half-wave plate; QPR, quadrant photoreceiver;  PBS, polarizing beamsplitter; CoM, center of mass.}
		\label{on_axis_concept}
	\end{figure*}
	
	The concept of the inter-spacecraft laser interferometry for geodesy in an on-axis configuration is shown in Fig.~\ref {on_axis_concept}. The twin spacecraft, operating in a trailing formation, orbit at altitudes 300–500 km above Earth with an approximate separation of 200 km in a near-polar orbit. In the reference spacecraft, the laser is frequency-stabilized to an optical cavity via the Pound–Drever–Hall (PDH) technique \cite{drever_laser_1983}. To operate in a transponder scheme, the laser in the transponder is frequency-offset-locked to the received signal from the reference spacecraft. The identical optical benches in the two spacecraft utilize polarizing optics to enable monoaxial transmission and reception of laser beams. The combination of a mirror (M1) and a quarter-wave plate (QWP) converts the perpendicular-polarized (S-pol) beam reflected by the polarizing beamsplitter (PBS) into the parallel-polarized(P-pol) light, thereby directing the majority of the laser power toward the remote spacecraft after retransmission through the PBS. In contrast to the off-axis configuration, the TX and RX reference points, which define the pivot points to achieve the minimal TTL coupling for the TX beam and the RX beam \cite{Muller2017PhD}, are co-located in the aperture of the optical bench. The fast steering mirror (FSM), the aperture, and the quadrant photoreceivers (QPRs) are arranged to establish optical conjugate planes using imaging systems, making the path length insensitive to the spacecraft's angular jitter. The DWS signals related to the spacecraft's pointing angle are fed back to the FSM or the attitude and orbit control system to maintain link strength and mitigate TTL coupling. 
    
\begin{figure}
		\centering
		\includegraphics[width=0.48\textwidth]{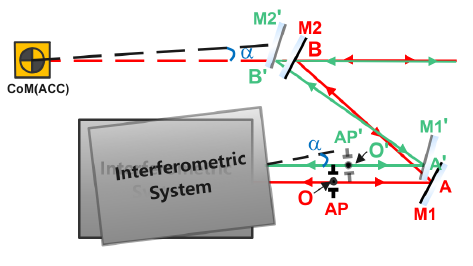}
		\caption{Schematic of reference point arrangements in an on-axis system using folding mirrors. The CoM is the virtual image of the AP, positioned via geometric unfolding of the optical path to serve as the center of the accelerator (ACC). The points of incidence on mirrors M1 and M2 are labeled A and B, respectively. The center of AP is designated as point O. The primed components M1$^{\prime}$, M2$^{\prime}$, AP$^{\prime}$, A$^{\prime}$, B$^{\prime}$  and O$^{\prime}$ denote the rotated configurations of M1, M2, AP, A, B and O, respectively, following a rotation of angle $\alpha$ about the CoM. The red and green traces illustrate the beam trajectories before and after rotation about the CoM. The total length $\text{OA+AB}$ equals the combined length $\text{O}^{\prime}\text{A}^{\prime} + \text{A}^{\prime}\text{B}^{\prime} + \text{B}^{\prime}\text{B}$, ensuring no piston effect during the rotation. M, mirror; AP, aperture; CoM, center of mass; ACC, Center of the Accelerator.}
		\label{reference_point_on_axis}
	\end{figure}

    Since an accelerometer can not be directly placed on the aperture of the optical bench, folding mirrors are implemented here to establish an effective reference point where the accelerometer is located, as explained in Fig.~\ref{reference_point_on_axis}. As the virtual image of the aperture, the effective reference point equivalent to the virtual vertex of the TMA in the off-axis configuration, coincides with the CoM of the spacecraft. As depicted in Fig.~\ref{reference_point_on_axis}, rotating the spacecraft about the center of mass (CoM) by an angle $\alpha$ induces an angular tilt in the RX beam while suppressing longitudinal phase shifts. Owing to the beam waist of the TX beam being virtually in the CoM, longitudinal phase shifts of the TX beam are suppressed as well when the spacecraft rotates about the CoM. As the mirror image of the aperture, the location of the CoM can be adjusted by repositioning the mirrors or altering their number, providing a high degree of flexibility to accommodate the accelerometer. 
    
	The off-axis configuration, characterized by a simplified optical bench and a high-precision TMA, does not utilize polarizing components. As evidenced by the GRACE-FO optical bench, this design incorporates a minimal number of critical components, reducing the stability requirements for the optical bench. However, its inherent racetrack configuration complicates telescope integration, thereby limiting the maximum measurable distance between spacecraft. In contrast, the on-axis topology leverages polarizing optics with a shared reference point on the optical bench, utilizing imaging systems and mirrors to achieve anti-parallel transmission and reception. This design minimizes the TTL coupling caused by angular jitter while eliminating the requirement for a TMA, a critical component in off-axis systems. The on-axis topology enables the utilization of a common telescope to enhance the received beam power and reduce the divergence of the TX beam, which is particularly beneficial for shot-noise-limited interferometers like the LISA mission \cite{LISA}. For example, in Fig.~\ref{on_axis_concept}, the telescope can be placed between the aperture and the folding mirrors.

    As discussed in \cite{yang2022axis}, the on-axis optical bench, using the same laser configuration and incorporating telescope imaging systems, shows an improvement in the $\mathrm{C/N_0}$, a key factor limiting the phase readout sensitivity, compared to the GRACE-FO optical bench. Moreover, the on-axis topology enables a more streamlined design, allowing for beam transmission and reception through a single telescope and baffle, which yields a more compact and space-efficient system. In contrast, off‑axis configurations such as the GRACE‑FO LRI use a retro‑reflector mounted separately from the optical bench, which increases assembly complexity and thermal‑control requirements. Because of its telescope adaptability, the on-axis topology offers distinct advantages for future potential space‑laser‑interferometry missions that require a longer interferometric arm.

	\section{METHODS}\label{methods}
	\subsection{Experimental setup}\label{setup}
	
	\begin{figure*}
		\centering
		\includegraphics[width=0.95\textwidth]{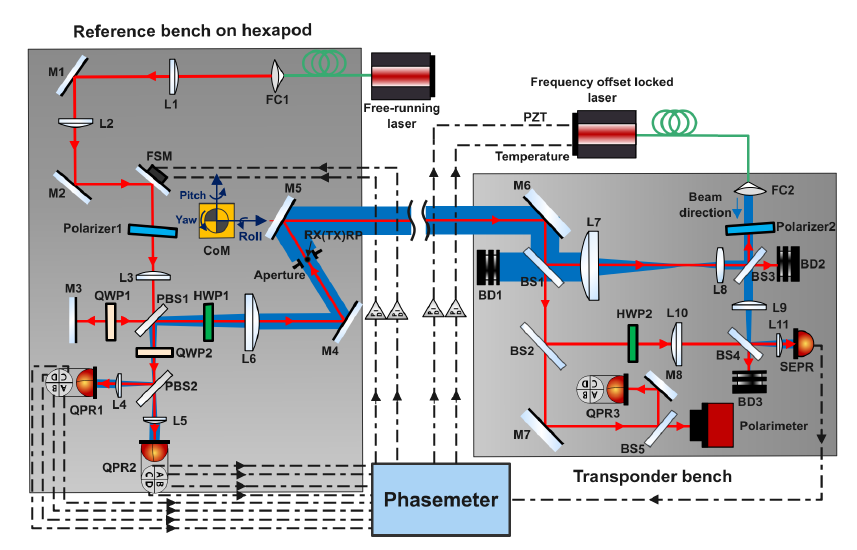}
		\caption{Schematic of the experimental setup. The red, blue, green, and black dashed lines denote the laser beam from the free-running laser, the laser beam from the frequency offset locked laser, the optical fiber, and the electronic connections, respectively. The coordinate system in the center of mass (CoM) is used to define yaw, pitch, and roll rotations. The definitions of the RX beam and the TX beam are referenced with the reference bench.  L, lens; M, mirror; FC, fiber collimator; FSM, fast steering mirror; QWP, quarter-wave plate; HWP, half-wave plate; SEPR, single-element photoreceiver; QPR, quadrant photoreceiver; BS, beamsplitter; PBS, polarizing beamsplitter; BD, beam dump; PZT, piezo-electric transducer.}
		\label{LRI_setup}
	\end{figure*}
	
\begin{figure*}
		\centering
		\includegraphics[width=0.95\textwidth]{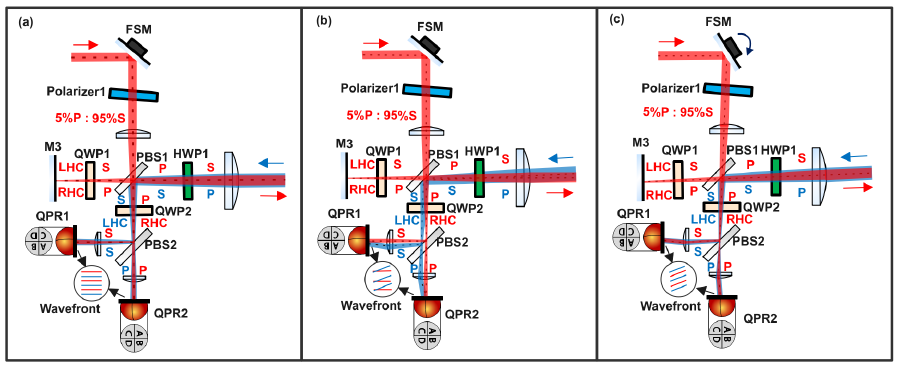}
		\caption{Beam configurations in the beam steering loop: (a) ideal alignment with zero DWS signal, (b) misalignment with non-zero DWS signal, and (c) realignment with zero DWS signal. The polarization states of the red and blue beams are denoted by red and blue text, respectively. The beam tilts illustrated in (b) and (c) are exaggerated for visualization, whereas in actual scenarios, the angular jitter would result in much smaller beam tilts. FSM, fast steering mirror; QWP, quarter-wave plate; HWP, half-wave plate; QPR, quadrant photoreceiver; PBS, polarizing beamsplitter; RHC, right-handed circularly polarized; LHC, left-handed circularly polarized; P, parallel-polarized; S, perpendicular-polarized.}
		\label{polarization_function_fsm}
	\end{figure*}

	Fig.~\ref{LRI_setup} shows the experimental setup, which consists of a reference bench, a transponder bench, and a phasemeter. As a modification of the concept we proposed in \cite{yang2022axis}, the reference bench is installed in a hexapod (Newport, HXP100-MECA), which provides six-degrees-of-freedom motions to simulate spacecraft attitude jitter in the experiment. In this experiment, the definitions of the RX beam and the TX beam are referenced with the reference bench. The transponder bench is fixed on an optical table and used to produce the RX beam, simulating the far-field beam from a remote spacecraft. The phasemeter tracks the phase of beat-note signals, which is then used to implement the active beam steering loops. The transponder scheme, validated in the GRACE-FO mission \cite{abich2019orbit,kornfeld2019grace,sheard2012intersatellite}, is implemented in this experiment, where the transponder laser is frequency-offset locked with the laser from the reference bench. As enlarged excerpts from the reference bench in Fig.~\ref{LRI_setup}, Fig.~\ref{polarization_function_fsm} illustrates the polarization states of beams and highlights the principle of the beam steering loop used in this experiment.

	On the reference bench, the laser is provided from a non-planar-ring-oscillator (NPRO) laser operating at a wavelength of 1064 nm. The phase sensitivity measurement is not addressed in this work, and thus the laser is operated in a free-running configuration, rather than the cavity-stabilized setup depicted in Fig.~\ref{on_axis_concept}. The laser light with about five mW enters the system via a single-mode polarization-maintaining (PM) fiber, and its beam waist is imaged onto an FSM (Physical Instrument, S-325) using two identical lenses L1 and L2 and two mirrors M1 and M2. The FSM is implemented here to adjust the TX beam in the beam steering loops. As illustrated in Fig.~\ref{polarization_function_fsm}, this process is facilitated by DWS signals, which are fed to the FSM to maintain the co-alignment between the RX beam and the TX beam. After the FSM, a linear polarizer is implemented to linearly polarize the incoming beam and to obtain a polarization ratio of 95\% S-pol to 5\% P-pol. The S-pol beam reflected by PBS1, with the most power, is then transmitted twice through a quarter-wave plate (QWP1), which flips its polarization state and sends it to the transponder bench as the TX beam. The remaining P-pol beam is directly transmitted through PBS1 and then interferes with the RX beam at PBS2, forming the local oscillator (LO).

	The RX beam is reflected off a mirror M5 and enters the system via a clipping at the aperture RX(TX)RP. By orienting the fast axis at 45$^{\circ}$, QWP2 transforms the P-pol LO beam and the S-pol RX beam into right-handed and left-handed circularly polarized beams, thereby enabling the heterodyne interference between the LO beam and the RX beam at PBS2. The beat-note beams in orthogonal polarization states are split by PBS2 and then detected by two InGaAs four-quadrant photoreceivers, QPR1 and QPR2, respectively. A half-wave plate (HWP1) is inserted after PBS1, with its fast axis oriented at 45$^{\circ}$, which converts both the TX and RX beams to S polarization.  Unlike the 45$^{\circ}$ linear polarization of the TX beam used in \cite{yang2022axis}, the configuration of pure S polarization of the TX beam prevents potential phase retardance caused by subsequent reflections, thereby eliminating the need for unpolarized coatings on subsequent mirrors. All waveplates, including QWP1, QWP2, and HWP1, are true zero-order waveplates, minimizing the TTL coupling due to phase retardance \cite{gu2017study}. 
	
	The FSM and the RX(TX)RP form a pair of conjugate images using a two-lens system of L3 and L6. This configuration magnifies the TX beam waist and images it onto the RX(TX)RP, thereby reducing the beam divergence angle. Simultaneously, the path length between the FSM and RX(TX)RP remains stable during TX beam deflection for angular jitter compensation, thus suppressing the TTL coupling. To mitigate effects from beam walks and diffraction at the RX(TX)RP, the imaging system composed of L4 and L3, and the imaging system composed of L4 and L6 simultaneously image the FSM and RX(TX)RP onto QPR1. For redundancy and balanced detection, L5 relays the images onto QPR2. A virtual CoM is defined on the bench, serving as the pivot point for the Hexapod's rotation in this experiment. The CoM is positioned as the mirror image of the RX(TX)RP.

\begin{figure}
		\centering
		\includegraphics[width=0.48\textwidth]{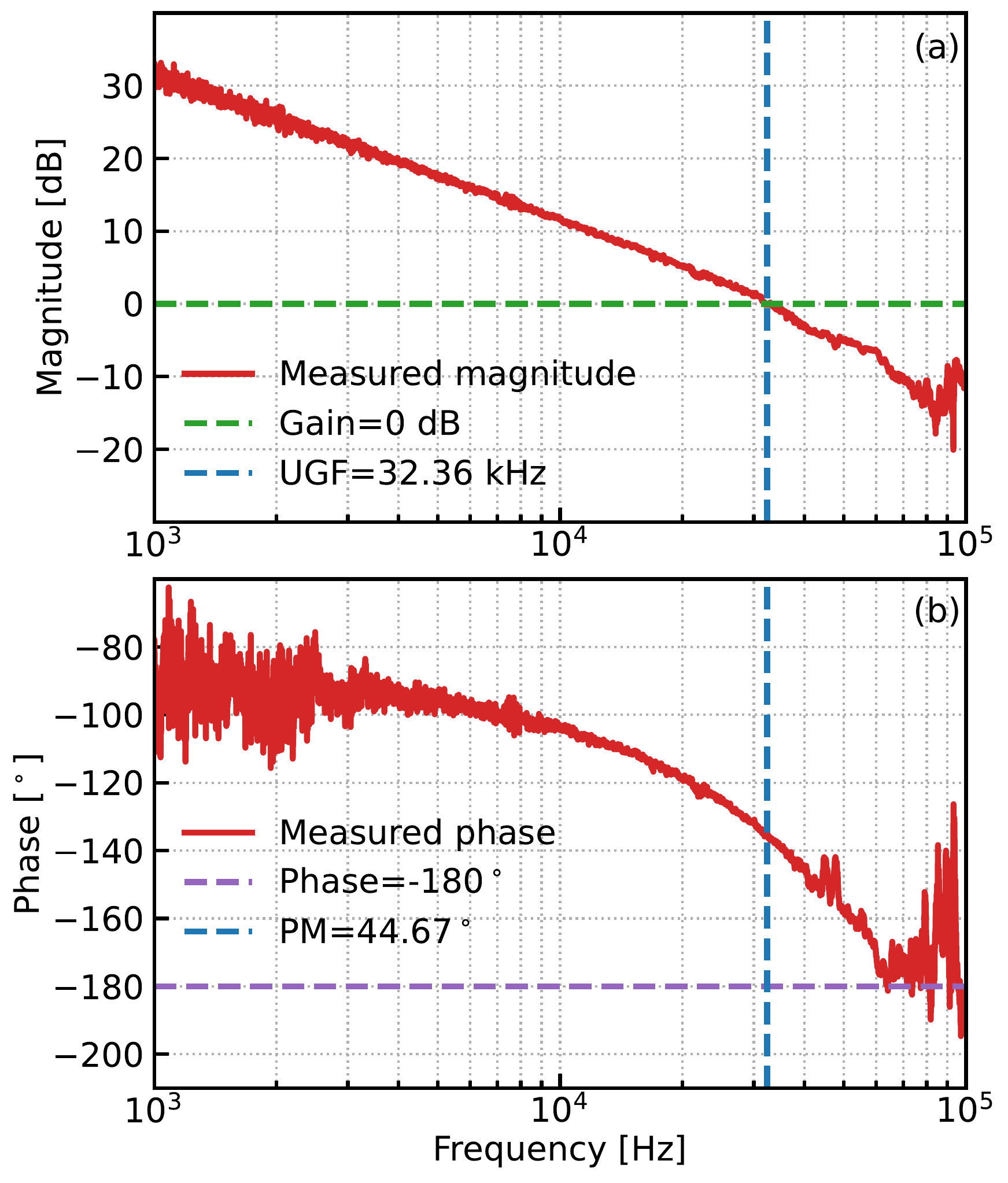}
		\caption{Measured open-loop transfer function of the laser frequency lock loop, including (a) the Bode magnitude plot and (b) the Bode phase plot. The laser frequency lock loop comprises a high-speed piezoelectric transducer (PZT) feedback loop and a low-speed temperature feedback loop. During the transfer function measurement, perturbations were injected into the PZT loop to characterize its frequency response. UGF, unity gain frequency; PM, phase margin.}
		\label{tf_laser_loop}
	\end{figure}

	On the transponder bench, a laser beam from a fiber collimator is split into two parts by a beam splitter BS3. The smaller portion is guided to the single-element photoreceiver (SEPR) as a reference beam, enabling  heterodyne detection with the TX beam from the reference bench. The phasemeter, employing all-digital phase-locked loops (ADPLLs), measures the beat-note frequency of the SEPR signal \cite{gerberding2013phasemeter, shaddock2006overview}. This frequency is fed back to a fast piezoelectric transducer (PZT) actuator and a slow temperature controller to regulate the laser frequency of the transponder laser \cite{shaddock2006overview, schwarze2018phase}, achieving a 7.3 MHz frequency offset lock. The open-loop transfer function of the laser frequency lock was measured using a frequency response analyzer (Moku:Lab, Liquid Instruments). As shown in Fig.~\ref{tf_laser_loop}, the laser frequency lock loop exhibits a unity-gain bandwidth of 32.36 kHz and a phase margin of 44.67$^{\circ}$.

\begin{figure}
		\centering
		\includegraphics[width=0.48\textwidth]{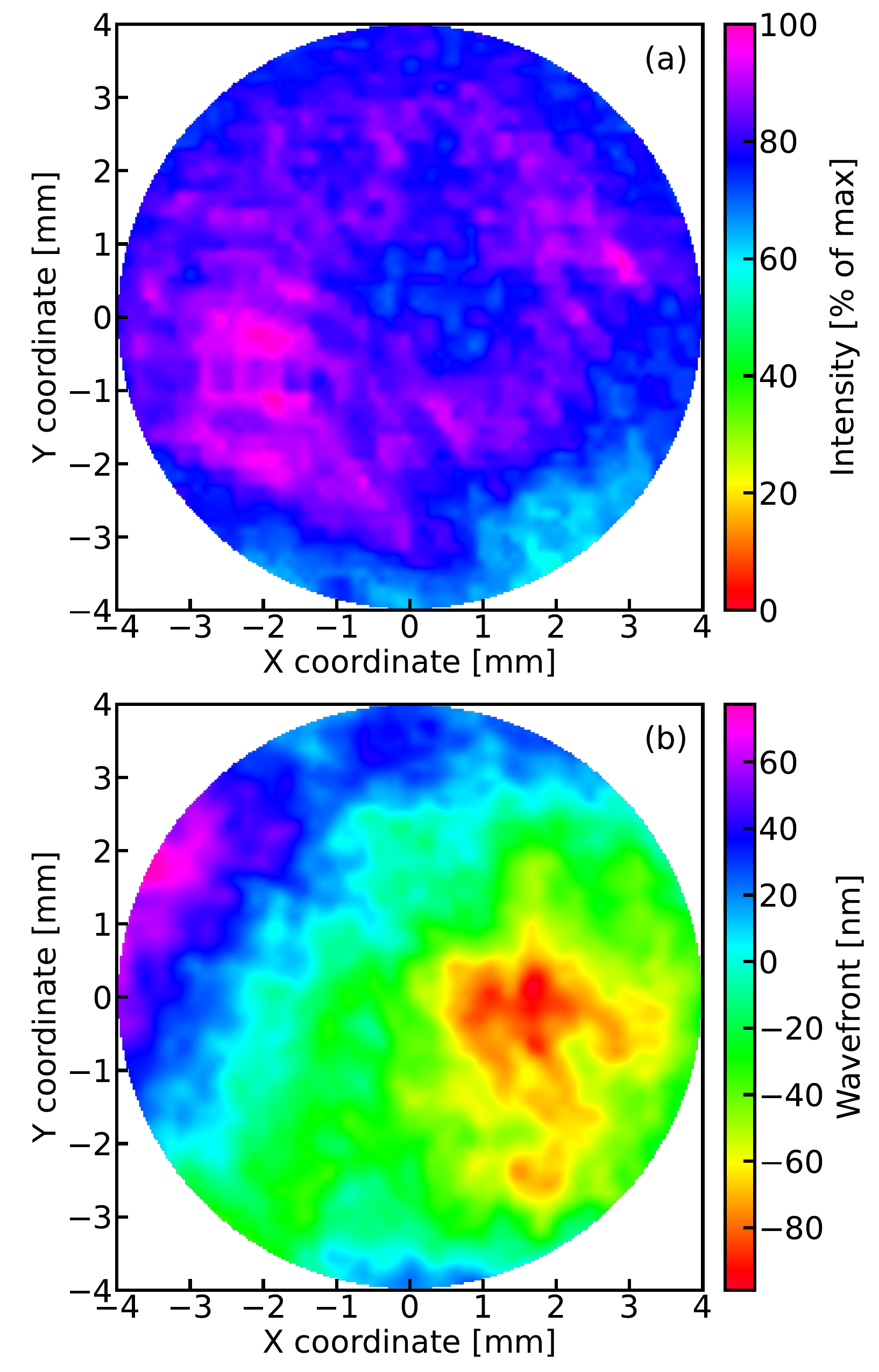}
		\caption{(a) Intensity profile and (b) wavefront profile of the RX Beam measured with a Shack–Hartmann wavefront sensor. The measurement result presented here is based on zonal wavefront reconstruction, and the wavefront tilts are corrected. The RX beam is clipped at the aperture, and thus only the central region of the beam distribution that falls within the aperture is presented in this analysis.}
		\label{shs_measurement}
	\end{figure}

	To simulate the RX beam for the reference bench, the beam reflected from BS3 is expanded using a two-lens system comprising L7 and L8, yielding a collimated beam with a $1/{e^2}$ radius of approximately 20.9 mm. This collimated beam, denoted as the RX beam, propagates to the reference bench. The wavefront and intensity profiles of the RX beam can be approximated as relatively flat within the central region, whereas the outer regions are clipped by the aperture, which has a radius of 4 mm. The beam distribution of the RX beam was characterized by a Shack-Hartmann wavefront sensor (Imagine Optics, HASO3-128-GE2) at a position 700 mm behind L7, equivalent to the aperture's location. The measured intensity and wavefront profiles are shown in Fig.~\ref{shs_measurement}. 
	The intensity distribution varies by less than 40\%, while the corresponding peak-to-valley (PTV) and root-mean-square (RMS) wavefront errors are 170 nm and 38 nm, respectively. 

    Although a beam with lower wavefront error and intensity drop could be achieved with increasing complexity of the system,  this was not considered necessary within the scope of this investigation. Even with a 40\% intensity drop in the RX beam, the $\mathrm{C/N_0}$ remained sufficient to ensure correct phase tracking by the phasemeter during this experiment. Throughout the experiment, only rotations of the reference bench around the CoM occurred, which induced a beam tilt at the RX (TX) RP. The corresponding longitudinal path length change resulting from this tilt constitutes the TTL coupling discussed in Section \ref{ttl}. While wavefront fluctuations in the RX beam introduced an initial offset in the DWS signal, this offset was calibrated using the method described in Section \ref{dws_dps}.
	
	The polarization state of the outgoing beam from the transponder bench is orthogonal to the polarization state of the TX beam, which is realized by rotating the linear polarizer polarizer2. Furthermore, HWP2 flips the polarization state of the TX beam, ensuring that it interferes with the beam from the transponder laser at BS4. A portion of the TX beam ends up on QPR3, where it is used to record the DPS signals that characterize the pointing of the TX beam. Additionally, a polarimeter (Thorlabs, PAX1000IR2) is installed on the transponder bench to monitor the polarization state of the TX beam.

	The phasemeter, originally developed for the LISA mission, is capable of tracking nanometer translations and nanoradian-scale tilts from the heterodyne signals of two laser beams.  The phase readout of the heterodyne signals is based on ADPLLs \cite{gerberding2013phasemeter,shaddock2006overview}, which are implemented in a field-programmable gate array (FPGA) platform. By combining the phase information from four individual channels, each connecting the signals from the segments of a QPR, the phasemeter calculates and processes the horizontal and vertical DWS signals \cite{heinzel2020tracking}. These DWS signals characterize the relative wavefront tilt of the two laser beams and are employed in the beam steering in this experiment.

	\subsection{DWS and DPS}\label{dws_dps}
    The DWS signals quantify angular misalignments by measuring the phase difference of the beat-note signal (e.g., left-right or top-bottom) via a QPR. This method enables precise determination of angular deviations between two beams, with a gain factor (DWS signal amplitude per wavefront tilt angle) ranging from $\sim 10^3$ rad/rad to $\sim 10^4$ rad/rad, depending on the interferometer’s design. More details about the DWS signal are included in Appendix~\ref{app_dws_dps}.

	The DWS signals acquired from QPR1 and QPR2 encode the relative wavefront tilt between the RX beam and the TX beam. The horizontal and vertical DWS signals are linearly proportional to the yaw and pitch misalignment over a small angular range \cite{wanner2012methods,heinzel2020tracking, morrison1994automatic}. With misalignment angles $\mathrm{\theta_{yaw}}$ (yaw) and $\mathrm{\theta_{pitch}}$ (pitch), the DWS signals are derived via the conversion matrix $T$, 
		\begin{equation}
		\begin{pmatrix}
			\mathrm{DWS_h}  \\  
			\mathrm{DWS_v} 
			\end{pmatrix} 
			= 
			\begin{pmatrix}
			T_{11}  &  T_{12}\\
			T_{21}  &  T_{22}
			\end{pmatrix}\begin{pmatrix}
			\mathrm{\theta_{yaw}}  \\  
			\mathrm{\theta_{pitch}} 
		\end{pmatrix}, 
		\label{def_matrix}
	\end{equation}
	\noindent
	where $\mathrm{DWS_h}$ and $\mathrm{DWS_v}$ denote the horizontal and vertical DWS signals, respectively. The diagonal elements $T_{11}$ and $T_{22}$ represent direct coupling factors between the misalignment angles and the corresponding DWS signals. The off-diagonal elements $T_{12}$ and $T_{21}$ describe cross-coupling effects, wherein yaw misalignment induces vertical DWS signals and pitch misalignment results in horizontal DWS signals. Intrinsic DWS offsets arising from inter-segment variations in wavefront errors, readout electronics, cable lengths, and QPR misalignment were observed in QPR-based measurements. These offsets were quantified by a modulated optical signal, which was artificially generated by amplitude-modulating the RX beam while blocking the LO beam. Throughout this paper, the measured DWS signals refer to the signals preprocessed by offset subtraction. 
	
	\begin{figure}
		\centering
		\includegraphics[width=0.48\textwidth]{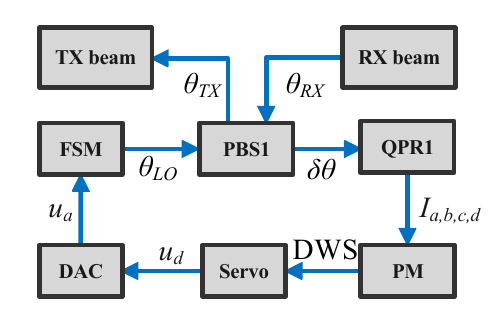}
		\caption{Block diagram of the beam steering loop. $\theta_{RX}$, incident angle of the RX beam on PBS1; $\theta_{TX}$, exit angle of the TX beam on PBS1; $\theta_{LO}$, incident angle of the LO beam on PBS1; $\delta\theta$, angle difference between $\theta_{LO}$ and $\theta_{RX}$; $I_{a,b,c,d}$, signal intensity at the segment (a, b, c, or d); $u_d$, digital signal; $u_a$, analog signal; FSM, fast steering mirror; QPR, quadrant photoreceiver; BS, beamsplitter; PM, phasemeter; DAC, Digital-to-analog converter; PBS, polarizing beamsplitter.}
		\label{DWS_loop}
	\end{figure}
	
	The desired direct coupling between the misalignment angles and the DWS signals dominates over parasitic cross-coupling effects. Therefore, two independent control loops utilize the horizontal and vertical DWS signals from QPR1 as servo signals to actuate the FSM and minimize the misalignment between the RX beam and the TX beam in this experiment. In the beam steering loop, the beam status of the LO beam, the TX beam, and the RX beam are illustrated in Fig.~\ref{polarization_function_fsm}, which shows the scenarios of ideal alignment with zero DWS signal, misalignment with non-zero DWS signal, and realignment with zero DWS signal.

	The block diagram of the beam steering loop is depicted in Fig.~\ref{DWS_loop}. The RX beam enters the optical bench with an incident angle $\theta_{RX}$ on PBS1. After polarization alignment, the RX beam interferes with the LO beam to produce heterodyne signals. The spacecraft attitude jitter introduces an angular misalignment $\delta\theta$ between the RX beam and the LO beam. This misalignment is detected by QPR1 and the phasemeter, which outputs a non-zero DWS signal. This signal is processed by a servo controller and converted to an analog voltage via a digital-to-analog converter (DAC), driving the FSM to adjust the LO beam and the TX beam. This feedback loop actively maintains $\delta\theta = 0$, minimizing the misalignment between the RX beam and the TX beam. This is the key property of both off-axis and on-axis schemes, ensuring that the outgoing beam points towards the remote spacecraft, independent of small misorientations of the local spacecraft. Although not included in the loop analysis depicted in Fig.~\ref{DWS_loop}, the waveplates and the imaging system are essential components of the overall loop. The waveplates determine the polarization states of the beams, thereby facilitating heterodyne interference at PBS2 and ensuring the transmission of the TX beam in the designated polarization state. Additionally, the imaging system suppresses the RX and LO beam walk at QPR1 and QPR2, thereby maintaining stable interferometric link operation within the loop.
	
	\begin{figure}[!htb]
		\centering
		\includegraphics[width=0.48\textwidth]{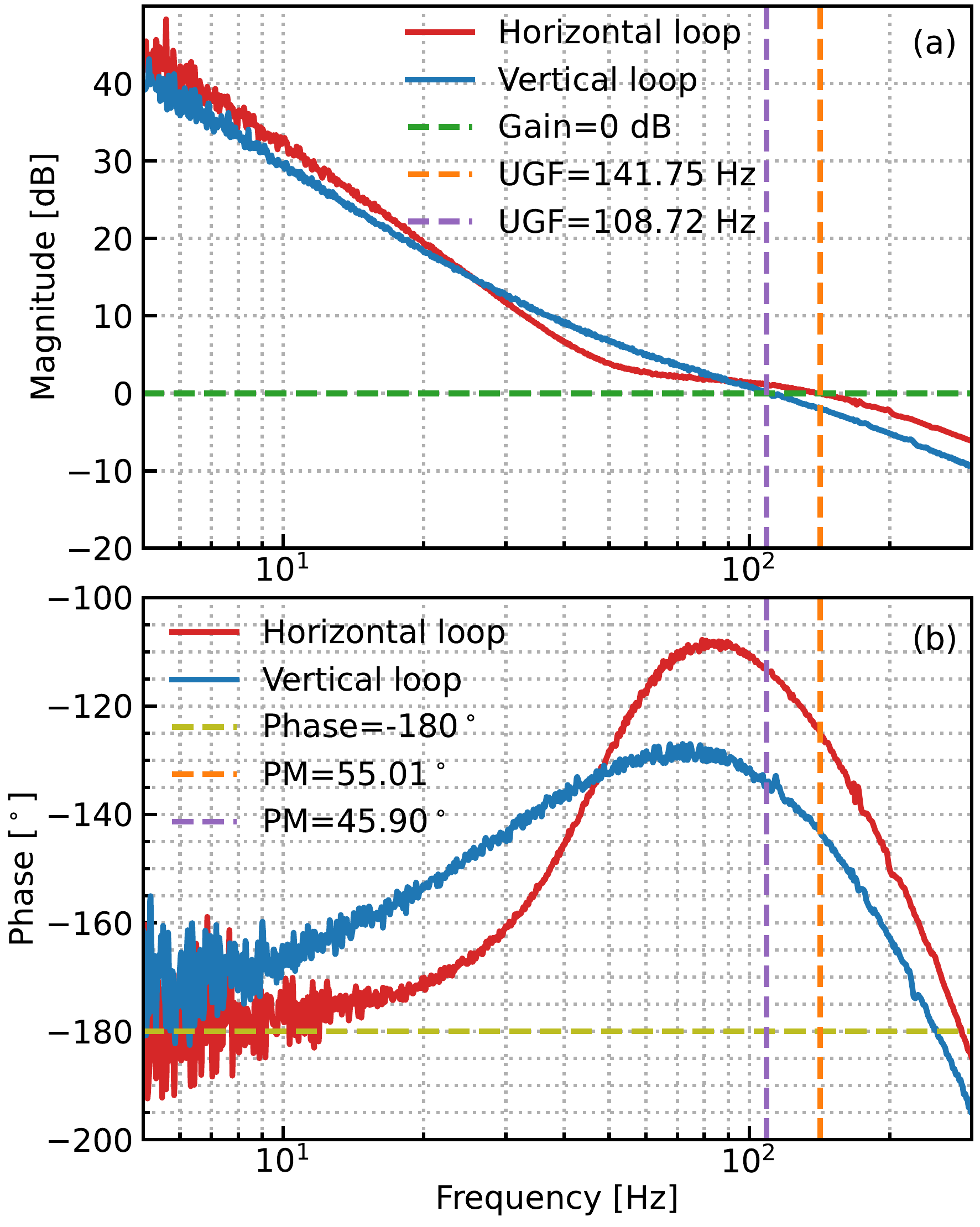}
		\caption{Measured open-loop transfer functions of the beam steering loops, including (a) the Bode magnitude plot and (b) the Bode phase plot. To introduce a perturbation, a swept sine signal was added to the signal $u_a$, as defined in Fig.~\ref{DWS_loop}, to measure the transfer function. The horizontal loop denotes the beam steering loop based on the horizontal DWS signal, and the vertical loop denotes the beam steering loop based on the vertical DWS signal. UGF, unity gain frequency; PM, phase margin.}
		\label{tf_DWS_loop}
	\end{figure}
	
	To assess the stability of the beam steering loops, the open-loop transfer functions were characterized using a frequency response analyzer (Moku:Lab, Liquid Instruments). As shown in Fig.~\ref{tf_DWS_loop}, the control loop based on the horizontal DWS signal exhibits a unity-gain bandwidth of 141.75 Hz and a phase margin of 55.01$^{\circ}$, whereas the control loop based on the vertical DWS signal demonstrates a unity-gain bandwidth of 108.72 Hz and a phase margin of 45.90$^{\circ}$. The bandwidths of both loops are much higher than the frequency of spacecraft attitude jitter, which is below 1 Hz in general \cite{daniel_intersatellite_2015}. Moreover, the phase margins of both loops exceed the typical threshold of 45$^{\circ}$, indicating that both beam steering loops can operate with adequate robustness.

	The DPS signal obtained from QPR3 is sensitive to the centroid shift of the TX beam and is therefore used to characterize the pointing of the TX beam. Similar to the DWS signal, the absolute pointing can be derived from the measured DPS signal using a conversion matrix that relates misalignment angles to DPS signals. The definition of the DPS signal is provided in Appendix~\ref{app_dws_dps}. The differences in responsivity between different segments of QPR3 can lead to unexpected errors when measuring the DPS signal. To mitigate this issue, the relative response differences between QPR3 segments were measured using homogeneous illumination from an integrating sphere. These response differences were then used to calibrate the true beam powers measured on QPR3 segments, ensuring accurate DPS signal determination.
	
	\section{RESULTS}
	\label{results}
	
	\subsection{Angle calibration}
	\label{ang_cal}
	
	\begin{figure*}[!htb]
		\centering
		\includegraphics[width=1.0\textwidth]{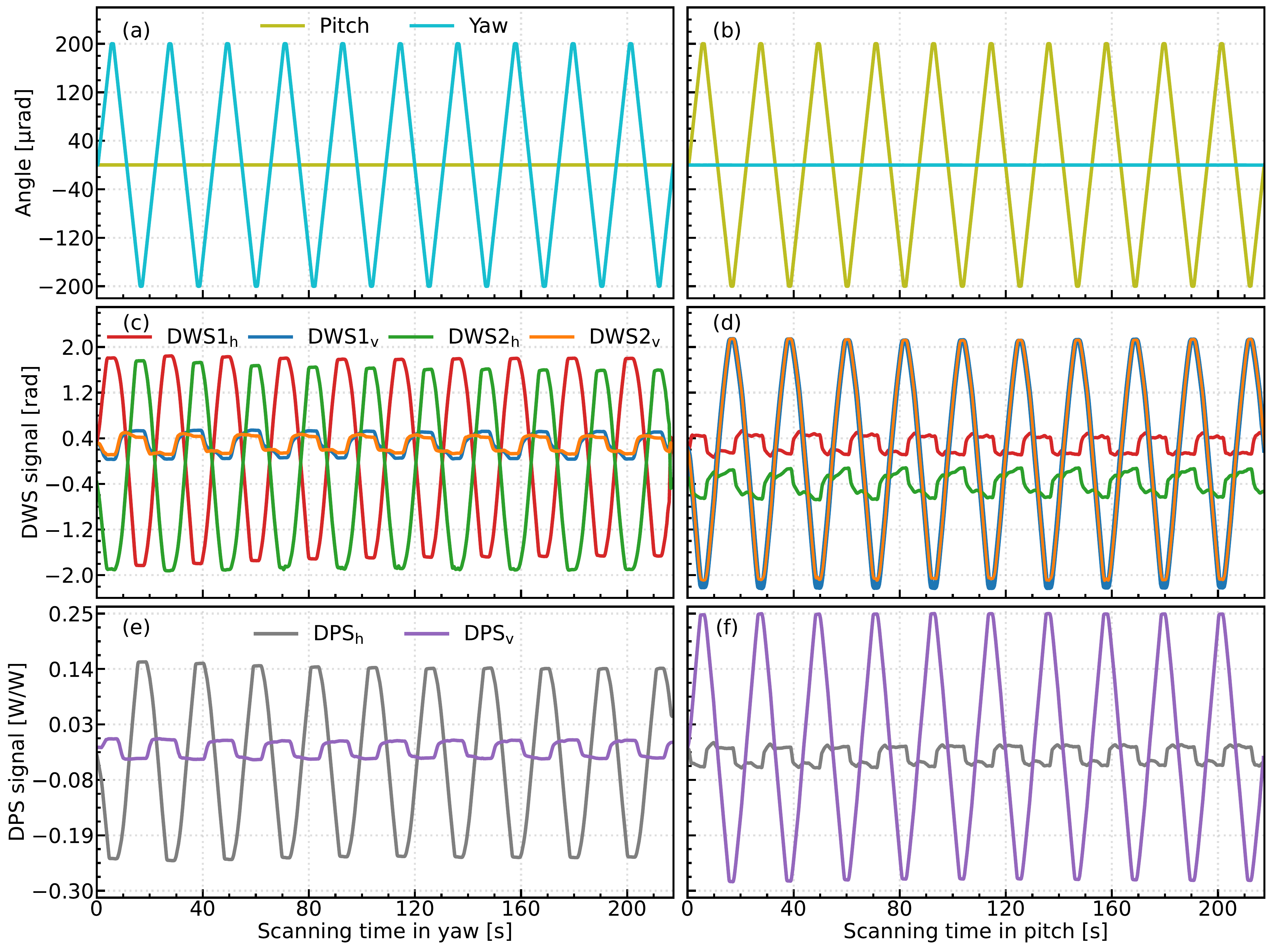}
		\caption{Angular scanning of the hexapod causes changes in the DWS and DPS signals in open-loop operation of the beam steering loops. (a, b) Rotation angles of the RX beam with respect to the TX beam as functions of the scanning time, monitored by the hexapod's position sensors. (c, d) Measured DWS signals as functions of the scanning time. (e, f) Measured DPS signals as functions of the scanning time. The angle information is provided from the hexapod's position sensors. The DWS1$\mathrm{_h}$ and DWS1$\mathrm{_v}$ denote the measured horizontal and vertical DWS signals from QPR1. The DWS2$\mathrm{_h}$ and DWS2$\mathrm{_v}$ denote the measured horizontal and vertical DWS signals from QPR2. The DPS$\mathrm{_h}$ and DPS$\mathrm{_v}$ denote the measured horizontal and vertical DPS signals from QPR3. }
		\label{dws_dps_calibration}
	\end{figure*}
	
	To determine the conversion matrices between the DWS signal and the wavefront tilt angle, and between the DPS signal and the TX pointing angle, the hexapod executed angular scans with a triangular wave pattern in the yaw and pitch directions, respectively. During the hexapod's motion, the beam steering loops were open. As depicted in Fig.~\ref{LRI_setup}, the yaw and pitch rotations are defined by the coordinate system in the CoM, and the rotating pivot point is located on the CoM.
	
	As shown in Fig.~\ref {dws_dps_calibration} (a) and (b), the rotation angle of the TX beam varies between -200 µrad and 200 µrad, as determined by the hexapod's position sensors. The DWS signals measured from QPR1 and QPR2, and the DPS signal measured from QPR3, are presented in Fig.~\ref{dws_dps_calibration} (c), (d), (e), and (f), respectively. The horizontal DWS and DPS signals exhibit a linear relationship with the rotation angle in the yaw direction, while the vertical DWS and DPS signals show a similar trend as the rotation angle of the pitch. A linear fit to the measured direct-coupling DWS and DPS signals as a function of rotation angle yields the corresponding coupling factors ($T_{11}$ and $T_{22}$), which are summarized in Table~\ref{conversion_matrix}.
	
	\begin{table}
		\caption{\label{conversion_matrix}%
			Conversion matrices between the measured signals and the rotation angles. Definition of \textrm{$T_{11}$}, \textrm{$T_{12}$}, \textrm{$T_{21}$}, and \textrm{$T_{22}$} is explained in Eq.~\ref{def_matrix}. The signals labeled DWS1, DWS2, and DPS in the table indicate that the matrices were derived from the measured DWS signals on QPR1 and QPR2, and the measured DPS signal on QPR3, respectively.
		}
		\begin{ruledtabular}
			\begin{tabular}{ccccc}
				\textrm{Signal}&
				\textrm{$T_{11}$} &
				\textrm{$T_{12}$}  &
				\textrm{$T_{21}$}  &
				\textrm{$T_{22}$}  \\
				\colrule
				DWS1 & 13666.62    &    -370.00   & -751.86  & -13243.35  \\ 
				DWS2 & -13818.17    &    -147.87  &  -31.13  & -13421.61   \\ 
				DPS  & -1386.71  & 24.53 & 33.15 & 1515.20       
			\end{tabular}
		\end{ruledtabular}
	\end{table}
	
	Residual alignment imperfections in the QPR introduce cross-coupling effects in the measurements. As shown in Fig.~\ref{dws_dps_calibration} (c), (d), (e), and (f), jumps in cross-coupling signals are observed when the hexapod reverses its scanning direction, while the cross-coupling signals remain relatively constant between these jumps. These jumps may result from the hysteresis and backlash effects \cite{NewportHXPManual}, which produce transient instability of the hexapod's motion during its settling time. Excluding the observed jumps, a linear fit of the measured cross-coupling DWS and DPS signals with respect to the rotation angle determines the cross-coupling factors ($T_{12}$ and $T_{21}$), as presented in Table~\ref{conversion_matrix}.

	\subsection{Beam Pointing}\label{pointing}
	
\begin{figure}
		\centering
		\includegraphics[width=0.48\textwidth]{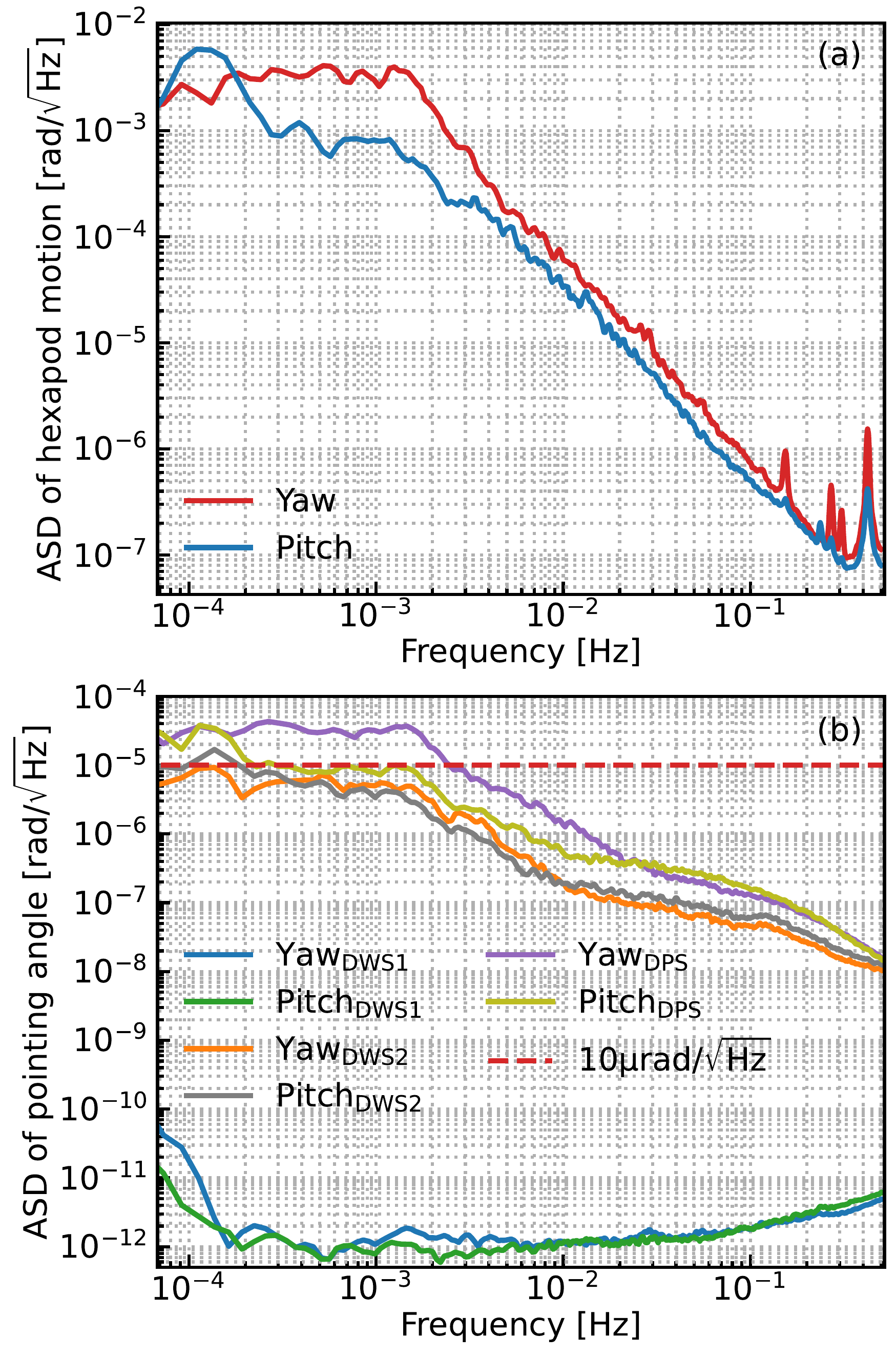}
		\caption{(a) Amplitude spectral density of the hexapod motion and (b) amplitude spectral density of the pointing angle as functions of frequency. The hexapod was commanded based on GRACE-FO jittering data from January 1st, 2019, while the beam steering loops are closed. The motion information used in (a) is obtained from the hexapod's position sensors. As shown in (b), all measured DWS and DPS signals are converted to the tilt angles using the conversion matrices obtained from section \ref{dws_dps}, and the subscripts indicate the signals used in the conversion. In-loop measurements obtained via DWS1, depicted by the blue and green lines in (b), show significantly lower values than the other pointing measurements.}
		\label{motion_pointing}
	\end{figure}

	The stability of the beam pointing in an LRI system affects the laser link, and the pointing noise can be coupled into ranging measurements through TTL coupling. To investigate the pointing stability of the on-axis architecture, the hexapod was commanded to perform typical motions derived from GRACE-FO pointing data recorded on January 1st, 2019 \cite{goswami2021analysis}, which artificially produce the TX beam pointing errors. The hexapod motion was updated at a frequency of 2 Hz over 15-hour continuous measurements, with the CoM as the nominal pivot point. The beam steering loops were closed during the hexapod motions. The coordinate information, provided from the hexapod's position sensors, was used to monitor hexapod motion in real time. Accordingly, the resulting amplitude spectral density (ASD) of the hexapod motion is shown in Fig.~\ref{motion_pointing} (a), which is consistent with the result in \cite{goswami2021analysis}. 
    Although several harmonics above 0.1 Hz are present, these frequency components are at the boundary of the GRACE-FO frequency band of 2 mHz to 0.1 Hz \cite{bachman2017flight} and thus do not affect our analysis here.
	
	To evaluate the co-alignment between the RX beam and the TX beam, the DWS signals from QPR1 and QPR2 were recorded. For comparison, the DPS signals from QPR3 were used to characterize the absolute TX beam pointing. Using the relation of Eq.~\ref{def_matrix}, the pointing angles were determined by multiplying the DWS and DPS signals by the inverse of the corresponding conversion matrices detailed in Table~\ref{conversion_matrix}. The corresponding ASD of the pointing angle is displayed in Fig.~\ref{motion_pointing} (b). In addition to the in-loop DWS measurements from the DWS1, out-of-loop DWS measurements from the DWS2 were performed as well. The horizontal (yaw) and vertical (pitch) co-alignments derived from the in-loop DWS measurements are below 10$^{-10}$ rad/$\mathrm{\sqrt{Hz}}$ at the frequency range from 0.07 mHz to 0.5 Hz, which indicates that the error signals are effectively suppressed by the control loops. For frequencies between 0.2 mHz and 0.1 Hz, both direction co-alignments derived from the out-of-loop DWS measurements are lower than 10 µrad/$\mathrm{\sqrt{Hz}}$, which fulfills the pointing requirement of NGGM mission \cite{nicklaus2020laser}. Compared to the pointing measurements of the GRACE-FO LRI \cite{schutze2014laser}, our measurements exhibit better pointing stability in the frequency range below 0.1 Hz.  
	
	The DPS measurements within the GRACE-FO frequency band (2 mHz to 0.1 Hz) reveal that the horizontal (yaw) pointing derived from the DPS is slightly above 10 µrad/$\mathrm{\sqrt{Hz}}$ between 2 mHz and 2.5 mHz, whereas the vertical (pitch) pointing derived from DPS is lower than 10 µrad/$\mathrm{\sqrt{Hz}}$ across the entire frequency band. In the experiment, all the optomechanical mounts are off-the-shelf components, and all the measurements were made in air, which inevitably leads to pointing jitters in the RX beam. Due to the jittering of the RX beam and the TX beam steering, the pointing angles derived from the DPS signals are higher than those derived from the DWS signals. In addition, the hexapod's parasitic motion in the lateral direction compromises the stability of the DPS. However, owing to the nearly flat-top distribution of the RX beam, the impact of this lateral motion on the DWS measurements is negligible. The measured DWS signals, which manifest the pointing of the TX beam with respect to the RX beam, which is the focus of our interest, not the absolute pointing of the TX beam derived from the measured DPS signals.

    Following the activation of the beam steering loops, the beam pointing stability in the yaw and pitch directions (Fig.~\ref{motion_pointing} (a)) was significantly enhanced, reaching the performance levels indicated by the orange and gray lines in Fig.~\ref{motion_pointing} (b). This represents a substantial improvement, particularly in the frequency range below 0.01 Hz where a two-order-of-magnitude reduction is observed. As the critical component in the beam steering loops, the noise performance of the FSM impacts the beam pointing stability. In the DWS2 measurements, thermal and air fluctuations limit further improvement at low frequencies. Additionally, although the imaging system maintains the optical path length almost consistent, the hexapod motion and the steering loops introduce  variations in the laser propagation path on the optical bench, further constraining the co‑alignment of the two beams in the low-frequency band.
	
	\subsection{Polarization}\label{polarization}
\begin{figure}
		\includegraphics[width=0.4\textwidth]{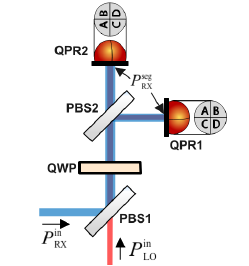}
		\caption{Schematic of the optical detection used in the on-axis LRI configuration. Unlike the reference bench shown in Fig.~\ref{LRI_setup}, the imaging systems are omitted here as their inclusion would not influence the polarization analysis. PBS, polarizing beamsplitter; QWP, quarter-wave plate; QPR, quadrant photoreceiver.}
		\label{diagram_pol}
	\end{figure}

	In the beam steering loop, the FSM actively deflects the beam to compensate for angular jitter-induced misalignment between the RX beam and the TX beam. The propagation of the deflected beam through the polarizing optics may cause a change in the polarization state of the TX beam. Fig.~\ref{diagram_pol} illustrates the optical detection scheme for the on-axis LRI. This polarization-dependent detection scheme, comprising PBS1, PBS2, QWP, and QPRs, converts deviations of the TX beam's polarization state from its nominal configuration (S polarization in this experiment) into a decline in the received RX beam power ($P_{\mathrm{RX}}^{\mathrm{seg}} $) at the remote spacecraft, ultimately leading to a reduction in the $\mathrm{C/N_0}$.

	As illustrated in Fig.~\ref{diagram_pol}, the LO beam and the RX beam impinge upon the detection system with input powers $P_{\mathrm{LO}}^{\mathrm{in}}$ and $P_{\mathrm{RX}}^{\mathrm{in}}$, respectively. Propagating the initial Stokes vectors through the optical system via Mueller calculus, we determine the polarization states at QPR1 and QPR2. The corresponding received powers are subsequently computed from the first Stokes parameter $\mathrm{S_0}$. Leveraging the condition $P_{\mathrm{RX}}^{\mathrm{in}} \ll P_{\mathrm{LO}}^{\mathrm{in}} $ characteristic of inter-spacecraft interferometric links \cite{Muller2017PhD}, the $\mathrm{C/N_0}$ is derived as  
	\begin{equation}
		\mathrm{C/N_0}\approx \frac{\eta P_{\mathrm{LO}}^{\mathrm{in}}P_{\mathrm{RX}}^{\mathrm{in}}\left(1-\mathrm{S_1}\right)R/64}{ \widetilde{\left[I_{\mathrm{PR}}\right]}+q_e R P_{\mathrm{LO}}^{\mathrm{in}}/4+\widetilde{\left[\delta P_{\mathrm{LO}}^{\mathrm{in}}/P_{\mathrm{LO}}^{\mathrm{in}}\right]}\left(RP_{\mathrm{LO}}^{\mathrm{in}}/8\right)^{2}},  
		\label{CNR_eq}
	\end{equation}
	\noindent
	where $\eta$ is the heterodyne efficiency, $R$ is the responsivity of the photodiode, $\mathrm{S_1}$ is the normalized second Stokes parameter describing the difference between P-pol and S-pol component in a beam,  $q_e$ is the electron charge, $\widetilde{\left[I_{\mathrm{PR}}\right]} $ denotes the power spectral density (PSD) of the photoreceiver's current, and $\widetilde{\left[\delta P_{\mathrm{LO}}/P_{\mathrm{LO}}\right]}$ denotes the PSD of the LO beam power fluctuation normalized to the average power. More details about the derivation of the $\mathrm{C/N_0}$ within the Mueller calculus framework are provided in Appendix~\ref{app_polarization}.

	During the pointing measurements described in Section \ref{pointing}, the polarization state of the TX beam was monitored by the polarimeter on the transponder bench, which recorded the Stokes parameters, azimuth angle, and ellipticity angle. In operational inter-spacecraft links, the TX beam propagates to the remote spacecraft and becomes the RX beam of the optical bench. Based on the polarization state measurements, we analyzed the impact of polarization state variations on the $\mathrm{C/N_0}$ using the relation of Eq.~\ref{CNR_eq}.

	To minimize potential phase retardance during measurements, the TX beam is transmitted through a non-polarizing optical path from the reference bench, situated after the M5, to the polarimeter, as shown in Fig.~\ref{LRI_setup}. Additionally, all waveplates are positioned between the two lenses of the imaging system to mitigate the polarization state fluctuation caused by beam deflection. This configuration ensures a consistent angle of incidence on the waveplates despite variations in the deflection angle of the TX beam induced by the FSM, thereby maintaining the polarization state of the beam.
	
	\begin{figure}
		\includegraphics[width=0.495\textwidth]{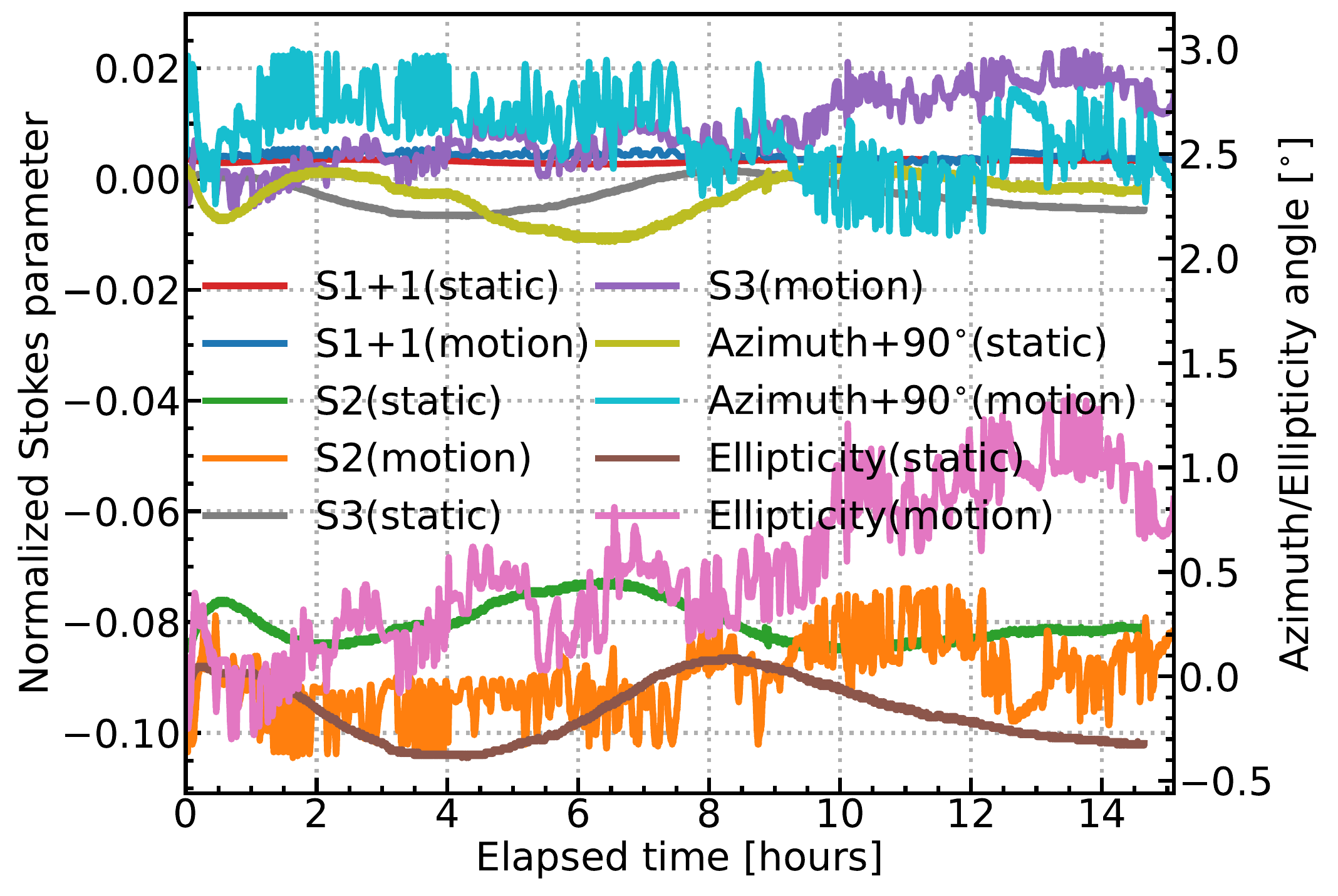}
		\caption{Temporal evolution of TX beam polarization state during hexapod pointing maneuvers (Fig.~\ref{motion_pointing} (a)) versus stationary operation.  The beam steering loops are closed during the hexapod pointing maneuvers.}
		\label{fig_pol}
	\end{figure}

	Fig.~\ref{fig_pol} shows the variation of the polarization state of the TX beam over 15 hours of continuous measurement. Complementing the pointing measurements described in Section \ref{pointing} (which incorporate hexapod motion), stationary hexapod measurements were taken as well for comparison. For all measurements, the degree of polarization (DoP) of the TX beam remained unity, indicating that depolarization did not occur during the measurements. The DoPs are quantified by the magnitude of the measured normalized Stokes vector: $\mathrm{DoP} = \sqrt{\mathrm{S_1^2}+\mathrm{S_2^2}+\mathrm{S_3^2}}$, where $\mathrm{S_i}$ denotes the normalized Stokes parameter. The azimuth angles remain confined to 90$^{\circ} \pm 0.025^{\circ}$ while ellipticity angles ($\varepsilon$) approach zero ($\left |\varepsilon  \right | < $ 0.11$^{\circ}$), confirming the TX beam's predominantly S-pol state. During stationary operation, the TX beam exhibited slow variation in Stokes parameters, azimuth angle, and ellipticity angle, which results from the ambient temperature fluctuations. Whereas the hexapod motion and the FSM beam steering produce rapid polarization-state variation, as validated in the pointing measurement results in Fig.~\ref{fig_pol}. During the pointing measurement, the Stokes parameter S1, azimuth angle, and ellipticity angle exhibited PTV fluctuations of 0.0027, 0.89$^{\circ}$, and 1.3$^{\circ}$, respectively. These variations correspond to a 0.14\% reduction in the $\mathrm{C/N_0}$ calculated via Eq.~\ref{CNR_eq}. Accounting for the contribution from the ambient temperature variation, the pure beam deflection contribution to $\mathrm{C/N_0}$ loss is expected to be less than the calculated values.

	\subsection{Tilt-to-length Coupling}\label{ttl}
	
	To investigate the TTL coupling of the optical bench, we drove the hexapod to rotate the entire reference bench around a fixed point, while keeping the transponder bench static. During the hexapod’s angular motion, the beam steering loops were closed so that the LO beam and the TX beam could be actively co-aligned with the RX beam. Then, the TTL coupling was characterized using the measured phases on QPD1, QPD2, and SEPR.

	The hexapod was programmed to execute 61-cycle angular scans in the yaw and pitch directions at frequencies of 0.49381 Hz and 0.38532 Hz, respectively. The rotational dynamics of the hexapod conform to a Jerk-limited SGamma profile \cite{NewportHXPManual}, which can be mathematically decomposed into a harmonic series within the fundamental mode of the scanning frequency. A detailed elaboration of this motion is provided in Appendix \ref{app_ttl}. The sampling frequency of phase measurement is 101.725 Hz, a multiple of the scanning frequency. During the measurements, the FSM's strain gauge sensors resolved angular displacements in yaw and pitch with a resolution of $\pm $0.05 µrad. In conjunction with the imaging system's magnification, the FSM's angular displacements were converted to the rotation angles of the setup, which were then used to determine the TTL coupling factor.

	The measurements were carried out in ambient air. To mitigate the influence of external disturbances such as air turbulence, thermal fluctuation, a dedicated Fast Fourier Transform (FFT) filter was implemented on the raw data derived from the measured phase at QPR1 and QPR2. The FFT filter retains only a specific number of harmonics in the spectral domain. The original spectral components of the hexapod's angular scans can be found in Appendix \ref{app_ttl}. The number of harmonics for the FFT filter was chosen to be 12, which is the trade-off between the measurement accuracy and the background noise. The filtered data were averaged over the period to obtain the round-trip longitudinal path length variations. 
	
	The linear TTL coupling factor approximates the positional displacement between the rotating pivot point and the CoM \cite{wegener2020tilt,wegener2022analysis}. By leveraging this correlation, the rotating pivot point can be adjusted according to the measured linear TTL coupling factor. The optimal pivot point was determined through iterative measurements until the positional mismatch was reduced below the hexapod's resolution threshold. The determined optimal pivot point was implemented in the final TTL measurements.
	\begin{figure}
		\centering
		\includegraphics[width=0.48\textwidth]{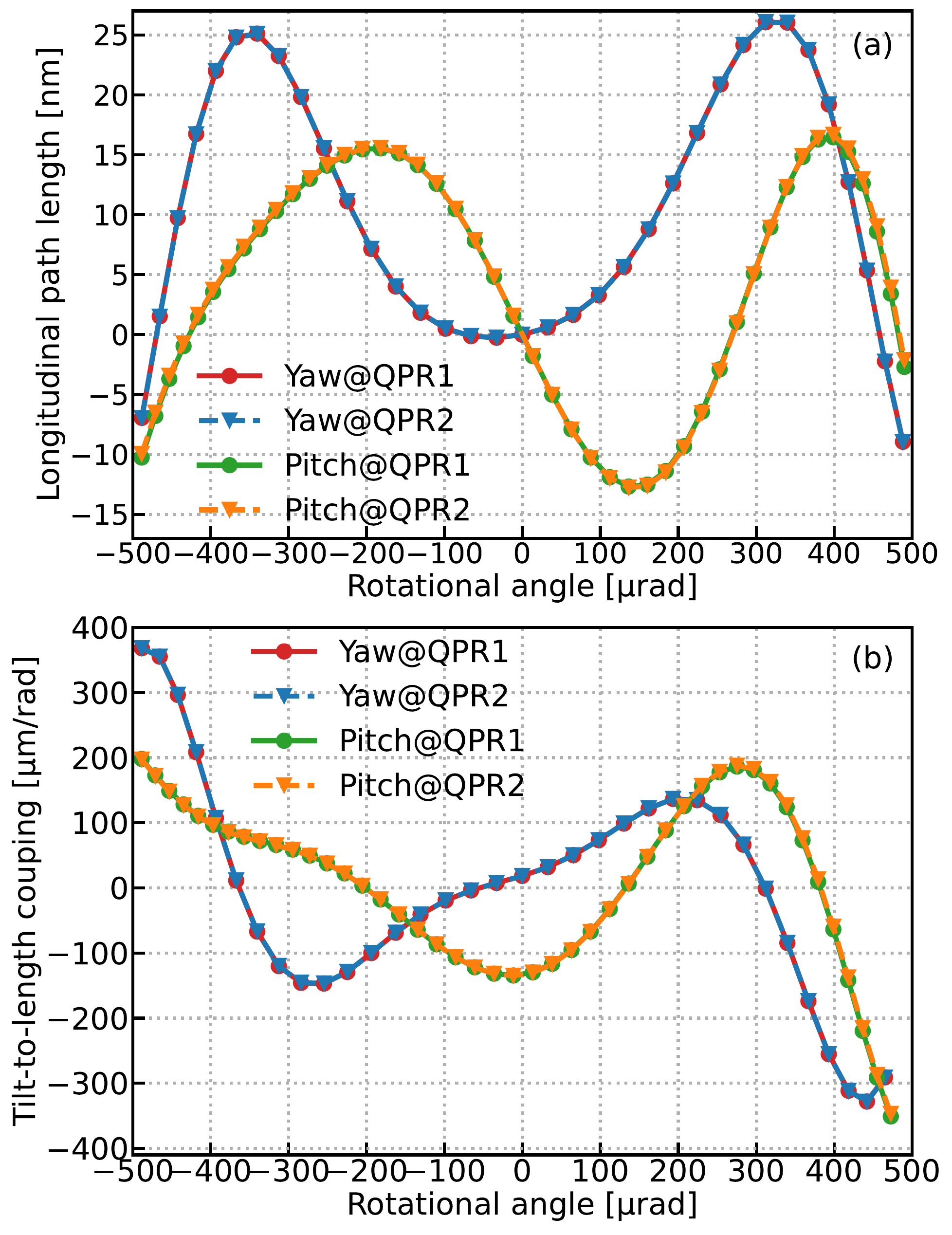}
		\caption{(a) Round-trip longitudinal path length variations and (b) tilt-to-length coupling factor as functions of rotation angle. The above results include 12 harmonics obtained by FFT filtering.}
		\label{lps_ttl}
	\end{figure}
	
	The longitudinal path length variations and TTL coupling factor with $\pm$500 µrad rotation angle in yaw and pitch directions are shown in Fig.~\ref{lps_ttl} (a) and (b), respectively. The QPR1 measurements closely match the QPR2 results, demonstrating the effectiveness of the balanced detection scheme as a redundant design. The longitudinal path length variations in the yaw and pitch directions are constrained to magnitudes below 27 nm and 17 nm, respectively. Correspondingly, the measured TTL coupling values for these directions are less than 370 µm/rad and 350 µm/rad, respectively, exceeding the simulated results by roughly one order of magnitude \cite{yang2022axis}. 
    The GRACE-FO LRI requires an optical bench TTL coupling of less than 80 µm/rad within a range of $\pm$300 µrad \cite{nicklaus2017optical}, with similar requirements for the GRACE-C and NGGM missions. However, the measured TTL numbers within this range are 145 µm/rad (yaw) and 188 µm/rad (pitch), which remain approximately twice the specified limit.

	The measurement inaccuracies could result from several factors, including temperature fluctuations, imaging system imperfections, and breadboard-level setup instability. The TTL coupling arising from the optical bench and misalignment between the pivot point and the effective reference point are experimentally indistinguishable. Consequently, the genuine TTL coupling of the optical bench may be compromised during measurements. However, the prime source of the error stems from the precision limit of the hexapod motion. During the scanning process, unintended phase jumps arose during the directional reversals, leading to spectral leakage during the filtering stage. Additionally, instability in the pivot point during the scanning process resulted in unintended longitudinal path length fluctuations, which we define as longitudinal coupling. These mechanical and kinematic deficiencies introduce significant measurement inaccuracies.

    The longitudinal coupling can be estimated using Eq.~\ref{LPS_offset_first_order}. During $\pm$500 µrad rotations, pivot point position uncertainties of 4 µm in $xy$ directions and 2 µm in $z$ direction(as specified in \cite{NewportHXPManual}) resulted in measurement uncertainty of 9.74 nm and 4.0 nm in the longitudinal pathlength variation. These correspond to 36.1\% and 23.5\% of the measured magnitudes in yaw and pitch direction, respectively. Since the TTL coupling factor was determined by the difference between the measured longitudinal path length at adjacent angles, such significant uncertainty inevitably leads to substantial errors in the TTL coupling factor, as shown in Fig.~\ref{lps_ttl}(b). Further details regarding the estimation methodology are provided in the Appendix \ref{app_ttl}.

	To address the limitations of the hexapod, an auxiliary interferometry system must be implemented between the two benches to monitor the longitudinal coupling, which can be subtracted during the post-processing. This interferometry system requires an additional laser beam on the transponder bench to produce a stable heterodyne signal. Moreover, the interferometry system must share a common optical path with the TX beam, spanning from the RX(TX)RP to mirror M6. However, implementing such a configuration under the current setup is challenging due to spatial constraints and system complexity. To further improve the TTL measurements, we propose replacing the breadboard system with a robust engineering model and employing a hexapod with higher positional precision.

	\section{SUMMARY AND CONCLUSION}\label{summary}
	We present a laboratory-scale prototype that demonstrates the feasibility of an on-axis topology for future GRACE-like missions. The prototype consists of a transponder-based laser interferometric link established between two optical benches, with phase readout at a heterodyne signal of 7.3 MHz. The reference bench is mounted on a hexapod stage to emulate spacecraft attitude jitter, while the transponder bench is designed to reproduce the flat-top profile characteristic of beams from remote spacecraft. To suppress angular jitter-induced misalignment between the RX beam and the TX beam, two independent beam steering loops are implemented with bandwidths of 141.75 Hz and 108.72 Hz, and phase margins of 55.01$^{\circ}$ and 45.90$^{\circ}$, respectively.

	The calibration factors linking the DWS and DPS signals to the tilt angle were determined experimentally based on the dedicated hexapod scans. These calibrated factors were subsequently employed to assess the pointing stability of the system. Under angular jitter analogous to that of the GRACE-FO satellite, the measured pointing stability remained below 10 µrad/$\mathrm{\sqrt{Hz}}$ across the frequency range of 0.2 mHz to 0.5 Hz in both horizontal and vertical directions, meeting the GRACE-FO mission requirements. Additionally, the fluctuation of the TX beam's polarization state only contributes to a 0.14\% $\mathrm{C/N_0}$ degradation during the 15-hour pointing measurements. The TTL coupling was characterized in the laboratory over a tilt angle range of $\pm$500 µrad. The measured TTL values exceeded theoretical predictions, with the dominant limitation arising from positioning inaccuracies of the hexapod.
	
	The experimental results validate the principal feasibility of the proposed interferometer and highlight its potential as an alternative for future gravity recovery missions. The concepts and technologies developed herein can benefit future inter-spacecraft laser interferometry missions. To achieve higher precision in TTL measurements, it is essential to eliminate longitudinal coupling effects, enhance the hexapod's positioning accuracy, and improve the optical system's robustness. Due to the limitations of the breadboard model, phase sensitivity remains unaddressed in this work. Future experimental investigations of the phase noise of the on-axis LRI require characterization in a vacuum chamber using a robust engineering model.

	\begin{acknowledgments}

		The authors thank Yihao Yan and Laura M\"{u}ller for their helpful discussions. The authors acknowledge support by the Chinese Academy of Sciences (CAS) and the Max Planck Society (MPG) within the framework of the LEGACY cooperation on low-frequency gravitational wave astronomy (M.IF.A.~QOP18098). The authors also acknowledge support from the Deutsche Forschungsgemeinschaft (DFG, German Research Foundation) under Project-ID 434617780-SFB 1464, the Relativistic Geodesy 1128, and the Clusters of Excellence: Light and Matter at the Quantum Frontier-Foundations and Applications in Metrology (EXC2123, Project No.~390837967), and PhoenixD: Photonics, Optics, and Engineering-Innovation Across Disciplines (EXC2122, Project No.~390833453). Finally, the authors gratefully acknowledge support from  the German Aerospace Center (DLR) with funds from the Federal Ministry for Economic Affairs and Climate Action (BMWK), based on a decision of the German Bundestag (Grant No.~50OQ2301, building on Grants No.~50OQ0601, 50OQ1301, and 50OQ1801). K.Y.’s work is supported by NASA under award number 80GSFC24M0006.
		
	\end{acknowledgments}
	
	\appendix
	
	\section{DWS and DPS}\label{app_dws_dps}
	
	The DWS signals represent the phase differences between beams across the four segments of a QPR, enabling the measurement of relative wavefront tilt between two optical beams \cite{morrison1994automatic, wanner2012methods}. By detecting the phases $\varphi_{A}$, $\varphi_{B}$, $\varphi_{C}$, and $\varphi_{D}$ at the four segments (A, B, C, and D) of a QPR, the horizontal DWS signal $\mathrm{DWS_h}$ (difference between left and right) and the vertical DWS signal $\mathrm{DWS_v}$ (difference between top and bottom) can be calculated as
	\begin{equation}
		\begin{aligned}
			\mathrm{DWS_h}  & = \frac{\varphi_{A}+\varphi_{C}-\varphi_{B}-\varphi_{D} }{2}, \\
			\mathrm{DWS_v} & = \frac{\varphi_{A}+\varphi_{B}-\varphi_{C}-\varphi_{D} }{2}.
		\end{aligned}
		\label{DWS_definition}
	\end{equation}
	\noindent
	Each segment's phase is determined using an ADPLL \cite{gerberding2013phasemeter,shaddock2006overview}.
	
	The DPS signals describe the variations in received power among the segments of a QPR. These signals are sensitive to incoming beam shifts, making them helpful in characterizing beam pointing. Given the recorded powers $P_{A}$, $P_{B}$, $P_{C}$, and $P_{D}$ at the four segments (A, B, C and D) of a QPR, the horizontal DPS signal $\mathrm{DPS_h}$ (difference between left and right) and vertical DPS signal $\mathrm{DPS_v}$  (difference between top and bottom) can be expressed as
	\begin{equation}
		\begin{aligned}
			\mathrm{DPS_h}  & = \frac{P_{A}+P_{C}-P_{B}-P_{D} }{P_{A}+P_{C}+P_{B}+P_{D}}, \\
			\mathrm{DPS_v} & = \frac{P_{A}+P_{B}-P_{C}-P_{D} }{P_{A}+P_{C}+P_{B}+P_{D}}.
		\end{aligned}
		\label{DPS_definition}
	\end{equation}
	\noindent
	The DPS signals are the result of normalizing the total received power.

	\section{Polarization}\label{app_polarization}

    In the on-axis detection system depicted in Fig.~\ref{diagram_pol}, the LO beam and the RX beam enter with input powers $P_{\mathrm{LO}}^{\mathrm{in}}$ and $P_{\mathrm{RX}}^{\mathrm{in}}$, respectively. The P-pol LO beam is characterized by the normalized Stokes vector $\left(1,1,0,0\right)^T$. Through Mueller calculus analysis of the system including PBS1, QWP1, and PBS2, the Stokes vectors of the LO beams at QPR1 and QPR2 become$\left(1/2,-1/2,0,0\right)^T$ and $\left(1/2,1/2,0,0\right)^T$, respectively. Given that $\mathrm{S_0}$ (first element of Stokes vector) represents the total optical power of the beam, and assuming uniform illumination of the photoreceiver active area, the incident LO beam power on each segment $ P_{\mathrm{LO}}^{\mathrm{seg}} $ is given by
	\begin{equation}
		P_{\mathrm{LO}}^{\mathrm{seg}} = P_{\mathrm{LO}}^{\mathrm{in}}/8.
		\label{LO_seg_P}
	\end{equation}
	\noindent
	With an input normalized Stokes vector $\left(1, \mathrm{S_1},\mathrm{S_2},\mathrm{S_3}\right)^T$, the Stoke vectors of the RX beam at QPR1 and QPR2 can be derived by
	\begin{equation}
		\begin{aligned}
			\vec{S}_{QPR1} &= 1/4\left(1-\mathrm{S_1}, -1+\mathrm{S_1}, 0, 0\right)^T,\\
			\vec{S}_{QPR2} &= 1/4\left(1-\mathrm{S_1}, 1-\mathrm{S_1}, 0, 0\right)^T.
		\end{aligned}
		\label{stokes_rx}
	\end{equation}
	\noindent
	Accordingly, the incident RX beam power on each segment $P_{\mathrm{RX}}^{\mathrm{seg}}$ can be obtained by 
	\begin{equation}
		P_{\mathrm{RX}}^{\mathrm{seg}} = 1/16P_{\mathrm{RX}}^{\mathrm{in}} \left(1-\mathrm{S_1}\right).
		\label{RX_seg_P}
	\end{equation}
	\noindent
	Given that $ P_{\mathrm{LO}}^{\mathrm{seg}} \gg P_{\mathrm{RX}}^{\mathrm{seg}} $ in an inter-spacecraft interferometric link, the $\mathrm{C/N_0}$ can be approximated by the expression derived in \cite{Muller2017PhD}:
	\begin{equation}
		\mathrm{C/N_0}\approx\frac{2 \eta P_{\mathrm{LO}}^{\mathrm{seg}}P_{\mathrm{RX}}^{\mathrm{seg}}R}{ \widetilde{\left[I_{\mathrm{PR}}\right]}+2q_e R P_{\mathrm{LO}}^{\mathrm{seg}}+\widetilde{\left[\delta P_{\mathrm{LO}}^{\mathrm{seg}}/P_{\mathrm{LO}}^{\mathrm{seg}}\right]}\left(RP_{\mathrm{LO}}^{\mathrm{seg}}\right)^{2}  },  
		\label{CNR_ref}
	\end{equation}
	\noindent
	where $\eta$ is the heterodyne efficiency, $R$ is the responsivity of the photodiode, $q_e$ is the electron charge, $\widetilde{\left[I_{\mathrm{PR}}\right]} $ denotes the power spectral density (PSD) of the photoreceiver's current, and $\widetilde{\left[\delta P_{\mathrm{LO}}/P_{\mathrm{LO}}\right]}$ denotes the PSD of the received LO beam power fluctuation normalized to the average power. The three terms in the denominator ($\widetilde{\left[I_{\mathrm{PR}}\right]} $, $ 2q_e R P_{\mathrm{LO}}^{\mathrm{seg}}$, and$\widetilde{\left[\delta P_{\mathrm{LO}}^{\mathrm{seg}}/P_{\mathrm{LO}}^{\mathrm{seg}}\right]}\left(RP_{\mathrm{LO}}^{\mathrm{seg}}\right)^{2} $) of Eq.~\ref{CNR_ref} represent the noise of the photoreceiver, shot noise, and relative intensity noise of the laser, respectively. By substituting Eq.\ref{LO_seg_P} and Eq.\ref{RX_seg_P} into Eq.~\ref{CNR_ref}, we derive the polarization-dependent $\mathrm{C/N_0}$ for the interferometric link as
	\begin{equation}
		\mathrm{C/N_0}\approx \frac{\eta P_{\mathrm{LO}}^{\mathrm{in}}P_{\mathrm{RX}}^{\mathrm{in}}\left(1-\mathrm{S_1}\right)R/64}{ \widetilde{\left[I_{\mathrm{PR}}\right]}+q_e R P_{\mathrm{LO}}^{\mathrm{in}}/4+\widetilde{\left[\delta P_{\mathrm{LO}}^{\mathrm{in}}/P_{\mathrm{LO}}^{\mathrm{in}}\right]}\left(RP_{\mathrm{LO}}^{\mathrm{in}}/8\right)^{2}}. 
	\end{equation}
	\noindent

	\section{Tilt-to-length Coupling}\label{app_ttl}
	
	The hexapod is employed to generate periodic rotational motion for TTL measurements. Its rotational dynamics adhere to a jerk-limited SGamma profile, characterized by four parameters: maximum velocity, maximum acceleration, maximum jerk time, and minimum jerk time \cite{NewportHXPManual}. The rotation angle of the hexapod can be decomposed into a series of harmonics. Specifically, with the fundamental mode $f_0$, the rotation angle $Rot\left( t \right)$ at time $t$ can be expressed as
	\begin{equation}
		\begin{aligned}
			Rot\left( t \right)\approx \sum_{n=1}^{N}A_n \text{sin}\left( 2\pi f_0 nt \right),
		\end{aligned}	
		\label{motion_decom}
	\end{equation}
	\noindent
	where $N$ denotes the number of harmonics to include in the approximation, $n$ is the mode number of the harmonic, and $A_n$ represents the amplitude at the nth order harmonic. By adjusting the maximum velocity, maximum acceleration, maximum jerk time, and minimum jerk time in the hexapod controller, the rotations of the hexapod were optimized to approximate a sinusoidal scanning profile, as quantified by enhancing the ratio of the fundamental mode in the rotation spectrum. The optimized rotations were employed in the TTL measurements.
	
	\begin{figure}
		\centering
		\includegraphics[width=0.48\textwidth]{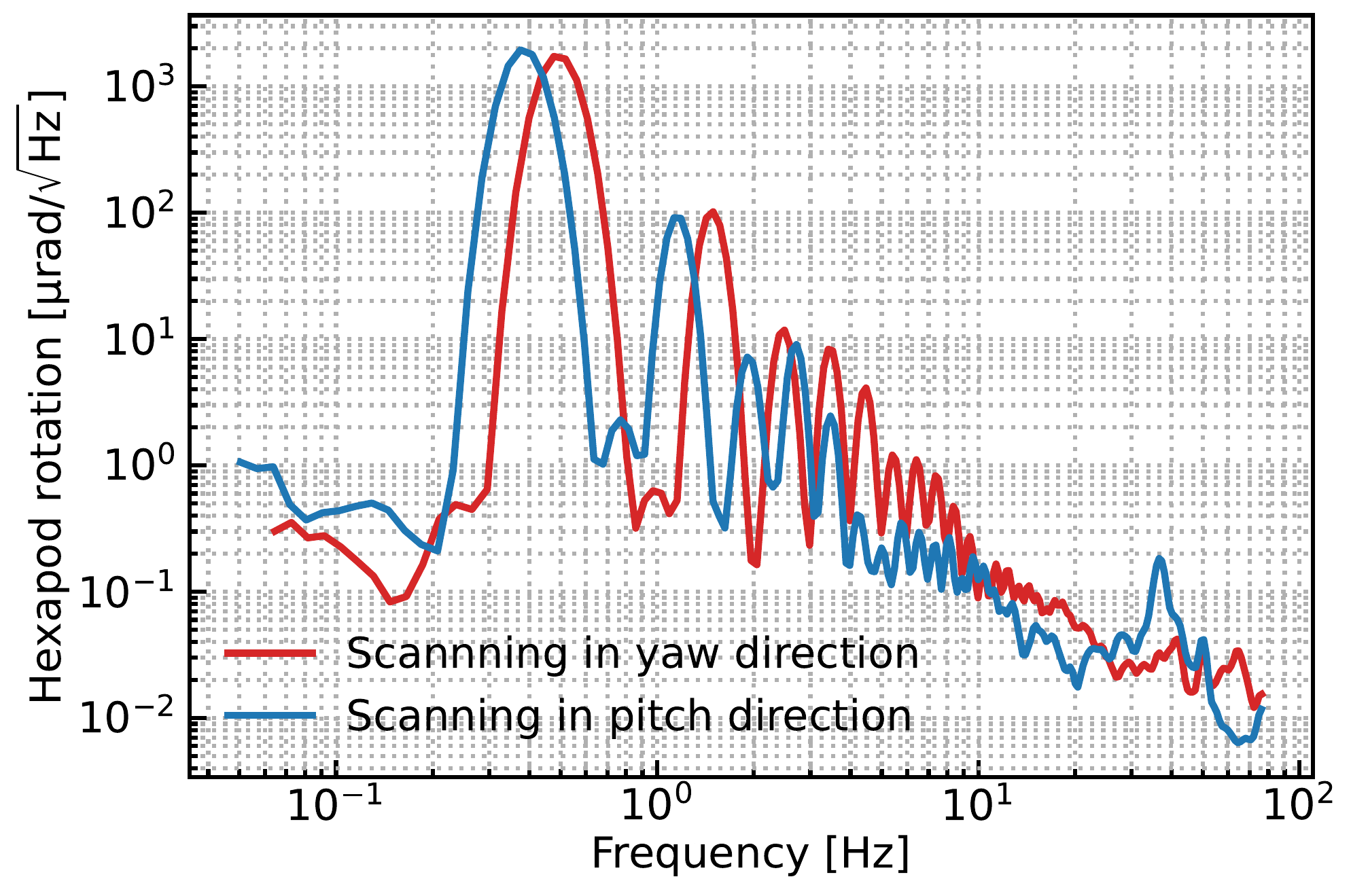}
		\caption{Hexapod angular displacement spectral density during tilt-to-length coupling characterization. This spectrum derives from 61-cycle angular scans, with rotation angles converted from FSM strain-gauge sensor measurements.}
		\label{move_asd}
	\end{figure}
    
    During the TTL measurement, the hexapod executed 61 cycles of angular scans in the yaw and pitch directions, respectively. With the beam steering loops closed, the FSM's angular displacements, measured via strain gauge sensors, were converted to equivalent hexapod rotation angles by multiplying the angular magnification of the imaging system. Using the measured angles, the resulting angular displacement spectral densities, computed via logarithmic frequency axis power spectral density (LPSD) \cite{trobs2006improved}, are presented in Fig.~\ref{move_asd}. Notably, our rotational scans revealed a series of harmonics in addition to the predominant fundamental mode.

    Similar to the off-axis LRI, the round-trip longitudinal pathlength variation that arises from the offset between the pivot point $\vec{p}_{pivot}$ and the reference point $\vec{p}_{RP}$  during the rotation is given by
   \begin{equation}
		\delta\rho = 2 R\cdot \left ( \vec{p}_{pivot} - \vec{p}_{RP}\right )\cdot \vec{e}_{LOS},  
		\label{LPS_offset}
	\end{equation}
	\noindent
    where $R$ defines the rotation matrix, and $\vec{e}_{LOS}$ denotes the unit vector along the line-of-sight direction (the laser propagation direction from the reference bench to the transponder bench) \cite{wegener2022analysis}. For small rotation angles $\theta_{yaw}$ and $\theta_{pitch}$,  a first‑order approximation yields
    \begin{equation}
		\delta\rho = 2\theta_{pitch} \Delta z-2\theta_{yaw} \Delta y, 
		\label{LPS_offset_first_order}
	\end{equation}
	\noindent
    with $\Delta y$ and $\Delta z$ denoting the mismatches in the hexapod's rotation coordinate frame \cite{schutze_retroreflector_2014}. Since the rotation frame used in the TTL measurements was derived from the hexapod's default coordinate system via a 14.4$^{\circ}$ yaw rotation transformation, this additional transfer matrix must be considered when estimating the longitudinal coupling. By applying Eq.~\ref{LPS_offset_first_order}, the positional errors from the specification and rotation angles, the potential longitudinal coupling can be determined.

	\nocite{*}
	
	\bibliography{apssamp}

@article{sheard2012intersatellite,
  title={Intersatellite laser ranging instrument for the {GRACE} follow-on mission},
  author={Sheard, BS and Heinzel, Gerhard and Danzmann, Karsten and Shaddock, DA and Klipstein, WM and Folkner, WM},
  journal={Journal of Geodesy},
  volume={86},
  pages={1083--1095},
  year={2012},
  publisher={Springer}
}

@article{tapley2004gravity,
	title={The gravity recovery and climate experiment: Mission overview and early results},
	author={Tapley, Byron D and Bettadpur, Srinivas and Watkins, Michael and Reigber, Christoph},
	journal={Geophysical research letters},
	volume={31},
	number={9},
	year={2004},
	publisher={Wiley Online Library}
}

@article{tapley2004grace,
	title={{GRACE} measurements of mass variability in the {Earth} system},
	author={Tapley, Byron D and Bettadpur, Srinivas and Ries, John C and Thompson, Paul F and Watkins, Michael M},
	journal={science},
	volume={305},
	number={5683},
	pages={503--505},
	year={2004},
	publisher={American Association for the Advancement of Science}
}

@article{nicklaus2020laser,
  title={Laser metrology concept consolidation for {NGGM}},
  author={Nicklaus, K and Cesare, S and Massotti, L and Bonino, L and Mottini, S and Pisani, M and Silvestrin, P},
  journal={CEAS Space Journal},
  volume={12},
  pages={313--330},
  year={2020},
  publisher={Springer}
}

@article{ghobadi2020grace,
	title={{GRACE Follow-On} laser ranging interferometer measurements uniquely distinguish short-wavelength gravitational perturbations},
	author={Ghobadi-Far, Khosro and Han, Shin-Chan and McCullough, Christopher M and Wiese, David N and Yuan, Dah-Ning and Landerer, Felix W and Sauber, Jeanne and Watkins, Michael M},
	journal={Geophysical Research Letters},
	volume={47},
	number={16},
	pages={e2020GL089445},
	year={2020},
	publisher={Wiley Online Library}
}

@article{haagmans2020esa,
	title={{ESA’s} next-generation gravity mission concepts},
	author={Haagmans, Roger and Siemes, Christian and Massotti, Luca and Carraz, Olivier and Silvestrin, Pierluigi},
	journal={Rendiconti Lincei. Scienze Fisiche e Naturali},
	volume={31},
	pages={15--25},
	year={2020},
	publisher={Springer}
}

@article{Kornfeld2019,
	doi = {10.2514/1.a34326},
	url = {https://doi.org/10.2514/1.a34326},
	year = {2019},
	month = may,
	publisher = {American Institute of Aeronautics and Astronautics ({AIAA})},
	volume = {56},
	number = {3},
	pages = {931--951},
	author = {Richard P. Kornfeld and Bradford W. Arnold and Michael A. Gross and Neil T. Dahya and William M. Klipstein and Peter F. Gath and Srinivas Bettadpur},
	title = {{GRACE}-{FO}: The Gravity Recovery and Climate Experiment Follow-On Mission},
	journal = {Journal of Spacecraft and Rockets}
}

@article{ward2014design,
  title={The design and construction of a prototype lateral-transfer retro-reflector for inter-satellite laser ranging},
  author={Ward, R L and Fleddermann, R and Francis, S and Mow-Lowry, C and Wuchenich, D and Elliot, M and Gilles, F and Herding, M and Nicklaus, K and Brown, J and Burke, J and Dligatch, S and Farrant, D and Green, K and Seckold, J and Blundell, M and Brister, R and Smith, C and Danzmann, K and Heinzel, G and Schütze, D and Sheard, B S and Klipstein, W and McClelland, D E and Shaddock, D A},
  journal={Classical and Quantum Gravity},
  volume={31},
  number={9},
  pages={095015},
  year={2014},
  publisher={IOP Publishing}
}

@article{tapley2005ggm02,
	title = {{GGM02} – {An} improved {Earth} gravity field model from {GRACE}},
	volume = {79},
	issn = {1432-1394},
	url = {https://doi.org/10.1007/s00190-005-0480-z},
	doi = {10.1007/s00190-005-0480-z},
	number = {8},
	urldate = {2025-06-02},
	journal = {Journal of Geodesy},
	author = {Tapley, B. and Ries, J. and Bettadpur, S. and Chambers, D. and Cheng, M. and Condi, F. and Gunter, B. and Kang, Z. and Nagel, P. and Pastor, R. and Pekker, T. and Poole, S. and Wang, F.},
	month = nov,
	year = {2005},
	pages = {467--478},
}

@article{yang2022axis,
  title={On-axis optical bench for laser ranging instruments in future gravity missions},
  author={Yang, Yichao and Yamamoto, Kohei and Dovale {\'A}lvarez, Miguel and Wei, Daikang and Esteban Delgado, Juan Jos{\'e} and M{\"u}ller, Vitali and Jia, Jianjun and Heinzel, Gerhard},
  journal={Sensors},
  volume={22},
  number={5},
  pages={2070},
  year={2022},
  publisher={MDPI}
}

@article{abich2019orbit,
	title = {In-{Orbit} {Performance} of the {GRACE} {Follow}-on {Laser} {Ranging} {Interferometer}},
	volume = {123},
	issn = {0031-9007, 1079-7114},
	url = {https://link.aps.org/doi/10.1103/PhysRevLett.123.031101},
	doi = {10.1103/PhysRevLett.123.031101},
	number = {3},
	urldate = {2024-09-26},
	journal = {Physical Review Letters},
	author = {Abich, Klaus and Abramovici, Alexander and Amparan, Bengie and Baatzsch, Andreas and Okihiro, Brian Bachman and Barr, David C. and Bize, Maxime P. and Bogan, Christina and Braxmaier, Claus and Burke, Michael J. and Clark, Ken C. and Dahl, Christian and Dahl, Katrin and Danzmann, Karsten and Davis, Mike A. and De Vine, Glenn and Dickson, Jeffrey A. and Dubovitsky, Serge and Eckardt, Andreas and Ester, Thomas and Barranco, Germán Fernández and Flatscher, Reinhold and Flechtner, Frank and Folkner, William M. and Francis, Samuel and Gilbert, Martin S. and Gilles, Frank and Gohlke, Martin and Grossard, Nicolas and Guenther, Burghardt and Hager, Philipp and Hauden, Jerome and Heine, Frank and Heinzel, Gerhard and Herding, Mark and Hinz, Martin and Howell, James and Katsumura, Mark and Kaufer, Marina and Klipstein, William and Koch, Alexander and Kruger, Micah and Larsen, Kameron and Lebeda, Anton and Lebeda, Arnold and Leikert, Thomas and Liebe, Carl Christian and Liu, Jehhal and Lobmeyer, Lynette and Mahrdt, Christoph and Mangoldt, Thomas and McKenzie, Kirk and Misfeldt, Malte and Morton, Phillip R. and Müller, Vitali and Murray, Alexander T. and Nguyen, Don J. and Nicklaus, Kolja and Pierce, Robert and Ravich, Joshua A. and Reavis, Gretchen and Reiche, Jens and Sanjuan, Josep and Schütze, Daniel and Seiter, Christoph and Shaddock, Daniel and Sheard, Benjamin and Sileo, Michael and Spero, Robert and Spiers, Gary and Stede, Gunnar and Stephens, Michelle and Sutton, Andrew and Trinh, Joseph and Voss, Kai and Wang, Duo and Wang, Rabi T. and Ware, Brent and Wegener, Henry and Windisch, Steve and Woodruff, Christopher and Zender, Bernd and Zimmermann, Marcus},
	month = jul,
	year = {2019},
	pages = {031101},
}

@article{kornfeld2019grace,
  title={{GRACE-FO}: the gravity recovery and climate experiment follow-on mission},
  author={Kornfeld, Richard P and Arnold, Bradford W and Gross, Michael A and Dahya, Neil T and Klipstein, William M and Gath, Peter F and Bettadpur, Srinivas},
  journal={Journal of spacecraft and rockets},
  volume={56},
  number={3},
  pages={931--951},
  year={2019},
  publisher={American Institute of Aeronautics and Astronautics}
}

@article{drever_laser_1983,
	title = {Laser phase and frequency stabilization using an optical resonator},
	volume = {31},
	issn = {1432-0649},
	url = {https://doi.org/10.1007/BF00702605},
	doi = {10.1007/BF00702605},
	number = {2},
	urldate = {2021-12-03},
	journal = {Applied Physics B},
	author = {Drever, R. W. P. and Hall, J. L. and Kowalski, F. V. and Hough, J. and Ford, G. M. and Munley, A. J. and Ward, H.},
	month = jun,
	year = {1983},
	pages = {97--105},
}

@phdthesis{Muller2017PhD,
	title={Design considerations for future geodesy missions and for space laser interferometry},
	author={M{\"u}ller, Vitali},
	year={2017},
	school={Hannover: Gottfried Wilhelm Leibniz Universit{\"a}t Hannover}
}

@article{schutze_retroreflector_2014,
	title = {Retroreflector for {GRACE} follow-on: {Vertex} vs point of minimal coupling},
	volume = {22},
	issn = {1094-4087},
	shorttitle = {Retroreflector for {GRACE} follow-on},
	url = {https://www.osapublishing.org/oe/abstract.cfm?uri=oe-22-8-9324},
	doi = {10.1364/OE.22.009324},
	number = {8},
	urldate = {2021-06-27},
	journal = {Optics Express},
	author = {Schütze, Daniel and Müller, Vitali and Stede, Gunnar and Sheard, Benjamin S. and Heinzel, Gerhard and Danzmann, Karsten and Sutton, Andrew J. and Shaddock, Daniel A.},
	month = apr,
	year = {2014},
	pages = {9324},
}

@article{bender2025short,
	title={Short-period mass variations and the next generation gravity mission},
	author={Bender, PL and Conklin, JW and Wiese, DN},
	journal={Journal of Geophysical Research: Solid Earth},
	volume={130},
	number={1},
	pages={e2024JB030290},
	year={2025},
	publisher={Wiley Online Library}
}

@article{gu2017study,
	title={Study of the retardance of a birefringent waveplate at tilt incidence by Mueller matrix ellipsometer},
	author={Gu, Honggang and Chen, Xiuguo and Zhang, Chuanwei and Jiang, Hao and Liu, Shiyuan},
	journal={Journal of Optics},
	volume={20},
	number={1},
	pages={015401},
	year={2017},
	publisher={IOP Publishing}
}

@article{gerberding2013phasemeter,
  title={Phasemeter core for intersatellite laser heterodyne interferometry: modelling, simulations and experiments},
  author={Gerberding, Oliver and Sheard, Benjamin and Bykov, Iouri and Kullmann, Joachim and Delgado, Juan Jose Esteban and Danzmann, Karsten and Heinzel, Gerhard},
  journal={Classical and Quantum Gravity},
  volume={30},
  number={23},
  pages={235029},
  year={2013},
  publisher={IOP Publishing}
}

@inproceedings{shaddock2006overview,
  title={Overview of the {LISA} {Phasemeter}},
  author={Shaddock, Daniel and Ware, Brent and Halverson, PG and Spero, RE and Klipstein, Bill},
  booktitle={AIP conference proceedings},
  volume={873},
  number={1},
  pages={654--660},
  year={2006},
  organization={American Institute of Physics}
}

@phdthesis{schwarze2018phase,
  title={Phase extraction for laser interferometry in space: phase readout schemes and optical testing},
  author={Schwarze, Thomas S},
  year={2018},
  school={Hannover: Gottfried Wilhelm Leibniz Universit{\"a}t Hannover}
}

@article{heinzel2020tracking,
  title={Tracking length and differential-wavefront-sensing signals from quadrant photodiodes in heterodyne interferometers with digital phase-locked-loop readout},
  author={Heinzel, Gerhard and {\'A}lvarez, Miguel Dovale and Pizzella, Alvise and Brause, Nils and Delgado, Juan Jos{\'e} Esteban},
  journal={Physical Review Applied},
  volume={14},
  number={5},
  pages={054013},
  year={2020},
  publisher={APS}
}

@article{schutze2014laser,
	title={Laser beam steering for GRACE Follow-On intersatellite interferometry},
	author={Sch{\"u}tze, Daniel and Stede, Gunnar and M{\"u}ller, Vitali and Gerberding, Oliver and Bandikova, Tamara and Sheard, Benjamin S and Heinzel, Gerhard and Danzmann, Karsten},
	journal={Optics express},
	volume={22},
	number={20},
	pages={24117--24132},
	year={2014},
	publisher={Optical Society of America}
}

@article{wanner2012methods,
  title={Methods for simulating the readout of lengths and angles in laser interferometers with Gaussian beams},
  author={Wanner, Gudrun and Heinzel, Gerhard and Kochkina, Evgenia and Mahrdt, Christoph and Sheard, Benjamin S and Schuster, S{\"o}nke and Danzmann, Karsten},
  journal={Optics communications},
  volume={285},
  number={24},
  pages={4831--4839},
  year={2012},
  publisher={Elsevier}
}

@article{goswami2021analysis,
	title={Analysis of GRACE follow-on laser ranging interferometer derived inter-satellite pointing angles},
	author={Goswami, Sujata and Francis, Samuel P and Bandikova, Tamara and Spero, Robert E},
	journal={IEEE Sensors Journal},
	volume={21},
	number={17},
	pages={19209--19221},
	year={2021},
	publisher={IEEE}
}

@article{nicklaus2022towards,
	title={Towards {NGGM}: {Laser} {Tracking} {Instrument} for the {Next} {Generation} of {Gravity} {Missions}},
	author={Nicklaus, Kolja and Voss, Kai and Feiri, Anne and Kaufer, Marina and Dahl, Christian and Herding, Mark and Curzadd, Bailey Allen and Baatzsch, Andreas and Flock, Johanna and Weller, Markus and Müller, Vitali and Heinzel, Gerhard and Misfeldt, Malte and Delgado, Juan Jose Esteban},
	journal={Remote Sensing},
	volume={14},
	number={16},
	pages={4089},
	year={2022},
	publisher={MDPI}
}

@inproceedings{landerer2024towards,
	title={Towards 30-years of mass change observations: {GRACE} {Follow}-{On} extended mission phase and {GRACE}-{Continuity} developments},
	author={Landerer, Felix W. and Wiese, David N. and Gross, Mike and Flechtner, Frank Michael and Save, Himanshu and Fischer, Sebastian and McCullough, Christopher M. and Dahle, Christoph and Bettadpur, Srinivas V. and Gaston, Robert and Snopek, Krzysztof},
	booktitle={AGU Fall Meeting Abstracts},
	volume={2024},
	pages={G21A--01},
	year={2024}
}

@article{wegener2020tilt,
	title={Tilt-to-length coupling in the {GRACE} {Follow-On} laser ranging interferometer},
	author={Wegener, Henry and M{\"u}ller, Vitali and Heinzel, Gerhard and Misfeldt, Malte},
	journal={Journal of Spacecraft and Rockets},
	volume={57},
	number={6},
	pages={1362--1372},
	year={2020},
	publisher={American Institute of Aeronautics and Astronautics}
}

@phdthesis{wegener2022analysis,
	title={Analysis of tilt-to-length coupling in the GRACE follow-on laser ranging interferometer},
	author={Wegener, Henry Paul},
	year={2022},
	school={Hannover: Gottfried Wilhelm Leibniz Universit{\"a}t Hannover}
}

@article{mandel2020architecture,
	title={Architecture and performance analysis of an optical metrology terminal for satellite-to-satellite laser ranging},
	author={Mandel, Oliver and Sell, Alexander and Chwalla, Michael and Schuldt, Thilo and Krauser, Jasper and Weise, Dennis and Braxmaier, Claus},
	journal={Applied Optics},
	volume={59},
	number={3},
	pages={653--661},
	year={2020},
	publisher={Optical Society of America}
}

@article{morrison1994automatic,
	title={Automatic alignment of optical interferometers},
	author={Morrison, Euan and Meers, Brian J and Robertson, David I and Ward, Henry},
	journal={Applied Optics},
	volume={33},
	number={22},
	pages={5041--5049},
	year={1994},
	publisher={Optical Society of America}
}

@manual{NewportHXPManual,
	author       = {{Newport Corporation}},
	title        = {HXP Controller User's Manual: Hexapod Motion Controller},
	organization = {Newport Corporation},
	address      = {Irvine, CA, USA},
	year         = {2015},  
	url          = {https://www.newport.com/medias/sys_master/images/images/h28/h28/8797113876510/HXP-Controller-User-Manual.pdf}
}

@misc{LISA,
Author = {Pau Amaro-Seoane and Heather Audley and Stanislav Babak and John Baker and Enrico Barausse and Peter Bender and Emanuele Berti and Pierre Binetruy and Michael Born and Daniele Bortoluzzi and Jordan Camp and Chiara Caprini and Vitor Cardoso and Monica Colpi and John Conklin and Neil Cornish and Curt Cutler and Karsten Danzmann and Rita Dolesi and Luigi Ferraioli and Valerio Ferroni and Ewan Fitzsimons and Jonathan Gair and Lluis Gesa Bote and Domenico Giardini and Ferran Gibert and Catia Grimani and Hubert Halloin and Gerhard Heinzel and Thomas Hertog and Martin Hewitson and Kelly Holley-Bockelmann and Daniel Hollington and Mauro Hueller and Henri Inchauspe and Philippe Jetzer and Nikos Karnesis and Christian Killow and Antoine Klein and Bill Klipstein and Natalia Korsakova and Shane L Larson and Jeffrey Livas and Ivan Lloro and Nary Man and Davor Mance and Joseph Martino and Ignacio Mateos and Kirk McKenzie and Sean T McWilliams and Cole Miller and Guido Mueller and Germano Nardini and Gijs Nelemans and Miquel Nofrarias and Antoine Petiteau and Paolo Pivato and Eric Plagnol and Ed Porter and Jens Reiche and David Robertson and Norna Robertson and Elena Rossi and Giuliana Russano and Bernard Schutz and Alberto Sesana and David Shoemaker and Jacob Slutsky and Carlos F. Sopuerta and Tim Sumner and Nicola Tamanini and Ira Thorpe and Michael Troebs and Michele Vallisneri and Alberto Vecchio and Daniele Vetrugno and Stefano Vitale and Marta Volonteri and Gudrun Wanner and Harry Ward and Peter Wass and William Weber and John Ziemer and Peter Zweifel},
Title = {Laser Interferometer Space Antenna},
Year = {2017},
Eprint = {arXiv:1702.00786},
}

@phdthesis{daniel_intersatellite_2015,
	address = {Hannover},
	title = {Intersatellite laser interferometry {Test} environments for {GRACE} {Follow}-{On}},
	school = {Leibniz University Hannover},
	author = {Daniel, Schütze},
	year = {2015},
}

@article{trobs2006improved,
  title={Improved spectrum estimation from digitized time series on a logarithmic frequency axis},
  author={Tr{\"o}bs, Michael and Heinzel, Gerhard},
  journal={Measurement},
  volume={39},
  number={2},
  pages={120--129},
  year={2006},
  publisher={Elsevier}
}

@article{heinzel2006lisa,
  title={LISA interferometry: recent developments},
  author={Heinzel, Gerhard and Braxmaier, Claus and Danzmann, Karsten and Gath, P and Hough, J and Jennrich, Oliver and Johann, Ulrich and R{\"u}diger, Albrecht and Sallusti, M and Schulte, H},
  journal={Classical and Quantum Gravity},
  volume={23},
  number={8},
  pages={S119},
  year={2006},
  publisher={IOP Publishing}
}

@inproceedings{nicklaus2017optical,
  title={Optical bench of the laser ranging interferometer on {GRACE Follow-On}},
  author={Nicklaus, K and Herding, M and Baatzsch, A and Dehne, M and Diekmann, C and Voss, K and Gilles, F and Guenther, B and Zender, B and Boehme, S and others},
  booktitle={International Conference on Space Optics—ICSO 2014},
  volume={10563},
  pages={738--746},
  year={2017},
  organization={SPIE}
}

@inproceedings{bachman2017flight,
  title={Flight phasemeter on the Laser Ranging Interferometer on the {GRACE Follow-On} mission},
  author={Bachman, B and De Vine, G and Dickson, J and Dubovitsky, S and Liu, J and Klipstein, W and McKenzie, K and Spero, R and Sutton, A and Ware, B and others},
  booktitle={Journal of Physics: Conference Series},
  volume={840},
  number={1},
  pages={012011},
  year={2017},
  organization={IOP Publishing}
}

@article{vallone2016interference,
  title={Interference at the single photon level along satellite-ground channels},
  author={Vallone, Giuseppe and Dequal, Daniele and Tomasin, Marco and Vedovato, Francesco and Schiavon, Matteo and Luceri, Vincenza and Bianco, Giuseppe and Villoresi, Paolo},
  journal={Physical review letters},
  volume={116},
  number={25},
  pages={253601},
  year={2016},
  publisher={APS}
}

@article{wu2024single,
  title={Single-photon interference over 8.4 km urban atmosphere: toward testing quantum effects in curved spacetime with photons},
  author={Wu, Hui-Nan and Li, Yu-Huai and Li, Bo and You, Xiang and Liu, Run-Ze and Ren, Ji-Gang and Yin, Juan and Lu, Chao-Yang and Cao, Yuan and Peng, Cheng-Zhi and others},
  journal={Physical Review Letters},
  volume={133},
  number={2},
  pages={020201},
  year={2024},
  publisher={APS}
}

@article{lu2022micius,
  title={Micius quantum experiments in space},
  author={Lu, Chao-Yang and Cao, Yuan and Peng, Cheng-Zhi and Pan, Jian-Wei},
  journal={Reviews of Modern Physics},
  volume={94},
  number={3},
  pages={035001},
  year={2022},
  publisher={APS}
}

@article{Luo2020,
doi = {10.1016/j.rinp.2019.102918},
url = {https://doi.org/10.1016/j.rinp.2019.102918},
year = {2020},
month = mar,
publisher = {Elsevier {BV}},
volume = {16},
pages = {102918},
author = {Ziren Luo and ZongKuan Guo and Gang Jin and Yueliang Wu and Wenrui Hu},
title = {A brief analysis to {Taiji}: {Science} and technology},
journal = {Results in Physics}
}

@article{Luo2016,
doi = {10.1088/0264-9381/33/3/035010},
url = {https://doi.org/10.1088/0264-9381/33/3/035010},
year = {2016},
month = jan,
publisher = {{IOP} Publishing},
volume = {33},
number = {3},
pages = {035010},
author = {Jun Luo and Li-Sheng Chen and Hui-Zong Duan and Yun-Gui Gong and Shoucun Hu and Jianghui Ji and Qi Liu and Jianwei Mei and Vadim Milyukov and Mikhail Sazhin and Cheng-Gang Shao and Viktor T Toth and Hai-Bo Tu and Yamin Wang and Yan Wang and Hsien-Chi Yeh and Ming-Sheng Zhan and Yonghe Zhang and Vladimir Zharov and Ze-Bing Zhou},
title = {{TianQin}: a space-borne gravitational wave detector},
journal = {Classical and Quantum Gravity}
}

@article{kawamura2011japanese,
  title={The {Japanese} space gravitational wave antenna: {DECIGO}},
  author={Kawamura, Seiji and Ando, Masaki and Seto, Naoki and Sato, Shuichi and Nakamura, Takashi and Tsubono, Kimio and Kanda, Nobuyuki and Tanaka, Takahiro and Yokoyama, Jun'ichi and Funaki, Ikkoh and others},
  journal={Classical and Quantum Gravity},
  volume={28},
  number={9},
  pages={094011},
  year={2011},
  publisher={IOP Publishing}
}
	
\end{document}